\documentclass[12pt]{iopart}
\usepackage[english]{babel}
\usepackage{amssymb,amsfonts,latexsym}

\usepackage{graphicx,epsfig,sidecap}
\usepackage{bm}% bold math

\usepackage{color}

\usepackage[colorlinks,%bookmarks=false,
citecolor=blue,linkcolor=red,urlcolor=blue]{hyperref}

\catcode`@=11 % If we need private TeX macros
\renewcommand\tableofcontents{%
  \section*{\contentsname}%
  \@starttoc{toc}%
}
\catcode`@=12% '@' is no more a character

\usepackage{simplewick}
\usepackage{cancel}

\def\be{\begin{equation}}
\def\ee{\end{equation}}

\def\bea{\begin{eqnarray}}
\def\eea{\end{eqnarray}}

\newcommand{\tw}{{\cal T}}
\def\t0{\tau_0}

\begin{document}

\title[Quantum quenches in 1+1D CFT]
{Quantum quenches in 1+1 dimensional conformal field theories}

\author{Pasquale Calabrese$^1$ and John Cardy$^{2,3}$}
\address{$^1$ SISSA and INFN, Via Bonomea 265, 
            34136 Trieste,  Italy.\\
         $^2$ Department of Physics, University of California, 
         Berkeley CA 94720, USA.\\
          $^3$ All Souls College, Oxford OX1 4AL, UK.}

\date{\today}

\begin{abstract}

We review the imaginary time path integral approach to the quench dynamics of conformal field theories. 
We show how this technique can be applied to the determination of the time dependence of correlation functions 
and entanglement entropy for both global and local quenches. 
We also briefly review other quench protocols. 
We carefully discuss the limits of applicability of these results to realistic models of condensed matter and cold atoms.

\end{abstract}

\maketitle

\tableofcontents

\section{Introduction}

The non-equilibrium quench dynamics of isolated quantum systems is a subject under intense theoretical investigation, 
triggered in part by recent experiments on trapped ultra-cold atomic 
gases \cite{uc,kww-06,tc-07,tetal-11,cetal-12,getal-11,shr-12,mmk-13,fse-13,langen-13,langen-15}.  
These cold gases are so weakly coupled to their environments as to allow the observation of essentially unitary time 
evolution over very long time scales. 
One of the central results which has emerged from these theoretical and experimental studies is that the dynamics of generic 
and integrable systems is dramatically different. Indeed, while generic systems (locally) 
thermalise (see \em e.g. \em\cite{d-91,s-94,rdo-08,bkl-10,bhc-10,bdkm-11,rf-11,rs-11,crfss-12,sks-14,gg-15,d-16}),
integrable models attain stationary values of local observables described by different statistical ensembles \cite{GGE}  
as a consequence of the infinity of constraints imposed by local and quasi-local integrals of motion.

One important point that has been clarified  in the last decade is
that the concepts of relaxation to a statistical ensemble and thermalisation are relative to subsystems
and not to the entire system (which, being initially in a pure state, will remain so for any time). 
Indeed, denoting the time evolved state of an infinite system as $|\psi(t)\rangle$, 
one introduces the reduced density matrix of the subsystem $A$ as $\rho_A(t)={\rm Tr}_{\overline A} |\psi(t)\rangle\langle \psi(t)|$ 
($\overline A$ being the complement of $A$).
If for an arbitrary finite subsystem $A$, 
$\rho_A(t)$ has an infinite time limit, then the system is said to admit (locally) a stationary state. 
If this long time limit equals the reduced density matrix of the Gibbs statistical ensemble then the system is said to thermalise.

The peculiar behaviour of integrable models is the main reason that motivated the development of 
advanced analytical tools to investigate their quench dynamics for a variety of different situations and realistic models,  
see \em e.g. \em Refs.~\cite{cc-06,gritsev,ce-13,m-13,a-12,ac-16,cd-16} and other reviews in this volume. 
However, many insights into the quench dynamics of many-body systems have come from the study of very simplified 
theories such as 1+1 dimensional conformal field theory (CFT), which has become an ideal playground in which to develop and check 
new ideas. 
For example, phenomena like the light-cone spreading of correlations \cite{cc-06,cc-07}, the linear increase of entanglement 
entropy \cite{cc-05}, 
and the structure of revivals in finite systems \cite{c-14} have been first described in CFT and later generalised to more 
realistic models, and even verified in experiments (see \cite{cetal-12} for the experimental measure of the light-cone 
spreading of correlations).  
Furthermore, once understood in CFT, some of these phenomena are readily generalised to more realistic models. 

The goal of this review is to give a self-contained presentation of the imaginary time path integral approach to 
the quench dynamics of CFTs. 
Many important features of integrable and generic systems will be mentioned only briefly. 
For a comprehensive treatment of all other aspects of the quench dynamics 
we refer the reader to the other excellent reviews in this volume and to the already 
existing ones in the literature \cite{silva,efg-14,dkpr-15,ge-15,a-16}.

The review is organised  as follows. In Sec.~\ref{sec:global} we review the imaginary time path integral approach to 
global quantum quenches and we apply it to CFT with a particular class of initial states. 
We describe the dynamical behaviour of correlation functions and entanglement entropy.
We highlight the phenomenon of light-cone spreading of entanglement and correlations.
We discuss in details the limit of applicability of these results to realistic models and finally we 
generalise the formalism to include more general classes of initial states, to finite systems and to perturbed CFTs. 
In Sec.~\ref{sec:local} we move to local quantum quenches and again describe the time evolution of entanglement and correlations.
We discuss finite size effects and how local quenches can be used to measure entanglement. 
In Sec.~\ref{sec:other} we briefly review other quench protocols that can be studied with the imaginary time formalism.

\section{CFT approach to global quantum quenches}
\label{sec:global}

The CFT approach to global quantum quenches was originally developed in \cite{cc-06,cc-07} 
(see also \cite{cc-05}) and later clarified and generalised in \cite{c-15}. 
Here we follow a presentation of the material which mixes these original references, with the 
goal of being pedagogical.

We consider the time evolution of a one-dimensional quantum system from an initial state $|\psi_0\rangle$, 
which we take to be translationally invariant, with short-range correlations and entanglement, \em e.g. \em
the ground state of a gapped hamiltonian $H_0$.  
At time $t=0$, a hamiltonian parameter is changed abruptly and, for times $t > 0$, 
the system evolves unitarily with a hamiltonian $H\neq H_0$. 
The time evolved state in the Schr\"odinger picture is clearly $e^{-i H t} |\psi_0\rangle$.
We are interested in the equal-time correlation functions of some local operators $\Phi_j(x_j)$
\be
\langle \psi_0| e^{i Ht} \Phi_1(x_1) \Phi_2(x_2)\dots \Phi_n(x_n) e^{-i H t} |\psi_0 \rangle.
\label{mult}
\ee
(The generalisation to correlations at different times is straightforward.) 
This expression can be rewritten in euclidean space (i.e. with imaginary time evolution) as 
\be
\langle \psi_0| e^{- H\tau_2} \Phi_1(x_1) \Phi_2(x_2)\dots \Phi_n(x_n) e^{- H \tau_1} |\psi_0 \rangle.
\label{mult2}
\ee 
This is nothing but the correlation function in an infinite euclidean strip of width $\tau_1+\tau_2$ and with
 boundary conditions on each edge corresponding to the state $|\psi_0\rangle$.
In order to recover the time evolution (\ref{mult}) from the strip geometry (\ref{mult2}), we should analytically continue 
the imaginary times as $\tau_1\to it$ and $\tau_2\to-it $. 
But, doing so, one would end up  with the nonsensical situation of a strip of zero total width.

In the case when the time evolution is governed by a 1+1 dimensional CFT, in \cite{cc-06,cc-07} we avoided this problem by appealing 
to the theory of boundary critical phenomena: the actual boundary conditions at $\tau=\tau_{1}$ and $\tau_2$ are replaced by conformal 
invariant boundary conditions $|\psi_0^* \rangle$ at $\tau= -\tau_0+\tau_1$ and $\tau=\tau_0+\tau_2$, 
where $\tau_0$ has been identified with the so-called extrapolation length. 
In the theory of boundary critical phenomena, this is justified on the basis of the renormalisation group (RG): 
the conformal invariant boundary condition corresponds to a fixed point of the RG, and $\tau_0$ measures the deviation of 
the actual state from this. 
For example, when dealing with a free massless boson,  possible conformal invariant boundary conditions are Dirichlet 
with zero or infinite field. 
Once we take a finite $\tau_0$, it is then possible to analytically continue $\tau_1=it$ and $\tau_2=-it $.

For simplicity in the calculations, it is convenient to translate in imaginary time by $\tau_0+\tau_1$, and so we consider the 
following correlation function in the euclidean strip with $\tau\in [0,2\tau_0]$
\be
\langle \psi_0^* | \Phi_1(x_1,\tau) \Phi_2(x_2,\tau)\dots \Phi_n(x_n,\tau)   |\psi_0^* \rangle\,,
\ee
where $\tau$ must be consider a real number during the course of the calculation and only at the end must be analytically continued 
to $\tau\to \tau_0+it$.

This argument is based on the assumption that the long-time behaviour after the quench should be insensitive to the details 
of the initial state as long as it has only short-range correlations. As has been argued in \cite{c-15}, at least for the case 
of evolution with a CFT hamiltonian, this is not in fact the case.
Again in \cite{c-15}  the prescription of \cite{cc-06,cc-07} was rephrased  in a way that its assumptions are clearer and 
so may be generalised. Indeed the above prescription is equivalent to assuming that the initial state has the form
\be
|\psi_0 \rangle\propto e^{-\tau_0 H} |\psi_0^*\rangle, 
\label{tau0B}
\ee
where $|\psi_0^*\rangle$ is the once again the conformally invariant boundary state. 
In this reinterpretation, it is clear also why it is not possible to take the limit $\tau_0\to0$, i.e. that  
the conformally invariant boundary states are not normalisable (as is well known) and the subsequent time evolution would  
not be well defined. 
The prescription (\ref{tau0B}) gives a finite result because it is equivalent to a smooth cutoff of 
the ultraviolet modes with energy larger than $\tau_0^{-1}$.
One could also interpret $\tau_0$ as being proportional to the correlation length (inverse mass) 
of the initial state since $|\psi_0^*\rangle$ has strictly vanishing correlations and the prefector $e^{-\tau_0 H}$
roughly correlates it on a scale $\tau_0$.

\subsection{Correlation functions following a global quench} 

We have at this point to deal with the calculation of correlation functions in a strip  $w=r+i\tau$  with $0<{\rm Im}\, w<2\tau_0$ 
in which at both edges ${\rm Im}\, w=0,2\t0$ there are the same conformal invariant boundary conditions $|\psi_0^*\rangle$.
This geometry can be mapped to the  the upper half-plane (UHP) ${\rm Im}\, z>0$ (with $|\psi_0^*\rangle$ at the boundary 
${\rm Im}\, w=0$) by the conformal transformation 
\be
w(z)=(2\tau_0/\pi)\ln z\,.
\label{logmap}
\ee
In the case where all fields $\Phi_i(w_i)$ are local primary  scalar
operators $\Phi_i(w_i)$, the expectation values in the strip and in the UHP are  related as
\be\label{eq:strip}
\langle\prod_i\Phi_i(w_i)\rangle_{\rm strip}
=\prod_iw'(z_i)^{-x_i/2} {\overline{w'(z_i)}}^{-x_i/2}\langle\prod_i\Phi_i(z_i)\rangle_{\rm UHP}\,,
\ee
where $x_i$ is the scaling dimension of the field $\Phi_i$. 
The asymptotic real time dependence follows from the analytic continuation
$\tau\to\tau_0+it$, and taking the limit $t,r_{ij}\gg \t0$. Note that after continuation $\bar z_i$ is no longer the complex conjugate of $z_i$, and the two factors on the rhs of (\ref{eq:strip}) must be continued separately.

\subsection{The one-point function of primary operator}
\label{1pt-glob}

In the UHP, the one-point function of a scalar primary field with scaling dimension $x$ is
$\langle\Phi(z)\rangle_{\rm UHP}= A_b^\Phi [2{\rm Im}(z)]^{-x}$. 
The normalisation factor $A^\Phi_b$ is a non-universal amplitude. 
In CFT the normalisations are chosen in such a way that $\langle\Phi(z_1) \Phi(z_2)\rangle_{\rm bulk}=|z_2-z_1|^{-2x}$. 
This choice fixes the amplitude $A^\Phi_b$ that turns out to depend both on 
the considered field $\Phi$ and on the boundary condition on the boundary $b$  \cite{cl-91}.
$A^\Phi_b$ vanishes if the expectation value of $\Phi$ on $|\psi_0^*\rangle$ vanishes, 
and thus $\langle\Phi(t)\rangle=0$, for all times.

When the primary field is not vanishing on the boundary,
performing the conformal mapping (\ref{logmap}) we obtain
\be
\langle \Phi(w)\rangle_{\rm strip}= |w'(z)|^{-x} \langle
\Phi(z(w))\rangle_{\rm UHP}= A^\Phi_b \left[\frac{\pi}{4\tau_0} 
\frac{1}{\sin(\pi\tau/(2\tau_0))}\right]^{x},
\label{phistrip}
\ee
that continued to real time $\tau=\t0+i t$ gives
\be
\langle\Phi(t)\rangle=
A^\Phi_b \left[\frac{\pi}{4\t0} \frac{1}{\cosh(\pi t/(2\t0))}\right]^{x}
\simeq A^\Phi_b \left(\frac{\pi}{2\t0}\right)^{x} e^{-x\pi t/2\t0}\,.
\label{onepoint}
\ee
Thus any primary field decays exponentially in time to zero (which is also the ground-state value), 
with a non-universal relaxation time $t_{\rm rel}^{\Phi}=2\tau_0/x \pi$. 
The ratio of the relaxation times of two different primaries equals 
the inverse of the ratio of their scaling dimensions and it is universal.

\subsection{The energy density}

An important exception to the exponential decay in time 
is the local energy density which corresponds to the $tt$ component of the
energy-momentum tensor $T_{\mu\nu}$. 
In CFT this is not a primary operator. Indeed, if it is normalised so that 
$\langle T_{\mu\nu}\rangle_{\rm UHP}=0$, in the strip we have \cite{BCN}
\be
\langle T_{tt}(r,\tau)\rangle=\frac{\pi c}{24(2\tau_0)^2}
\label{Ttt}
\ee
(where $c$ is the central charge  of the CFT) so that it does not decay in time.
Of course this is to be expected since the dynamics conserves
energy. A similar feature is expected to hold for other local
densities corresponding to globally conserved quantities which
commute with the hamiltonian, for example the magnetisation along $z$ in an anisotropic magnet with $U(1)$ symmetry in the 
$xy$ plane.

\subsection{Two-point function of primary operators}

Let us now consider the time evolution of a two-point function of primaries at distance $\ell$.
The two-point function in the upper half-plane assumes the general scaling form \cite{cardy-84} 
\be
\langle\Phi(z_1) \Phi(z_2) \rangle_{\rm UHP}=\left
(\frac{z_{1\bar2}z_{2\bar1}}{z_{12}z_{\bar1\bar2}z_{1\bar1}z_{2\bar2}}\right)^x
F(\eta)\,,
\label{2ptgen}
\ee
where $\eta={z_{1\bar1}z_{2\bar2}}/{z_{1\bar 2}z_{2\bar1}}$ is the four point ratio constructed with $z_1$, $z_2$ and 
their images $\bar z_1, \bar z_2$, 
the function $F(\eta)$ depends on the full operator content of the CFT. 
We need to map this four-point function to the strip and consider the images points at $w_1=0+i\tau$ and $w_2=\ell+i\tau$.
With some simple algebra, and continuing to $\tau=\tau_0+it$ (see
\cite{cc-05,cc-06,cc-07} for detailed calculations), we find for $t,\ell\gg\tau_0$
\begin{equation}
\fl
\langle \Phi(\ell, t) \Phi(0,t)\rangle \simeq  \left(\frac{\pi}{2\tau_0}\right)^{2 x}
\left(\frac{e^{\pi\ell/2\tau_0}+e^{-\pi\ell/2\tau_0}+2\cosh(\pi t/\tau_0)}
{(e^{\pi\ell/4\tau_0}-e^{-\pi\ell/4\tau_0})^2 \cosh^2(\pi
t/2\tau_0)}\right)^{x} F(\eta)\,.
\label{twott}
\end{equation}
After the conformal mapping and analytically
continuing the four-point ratio $\eta$ becomes for $\ell/\tau_0$ and
$t/\tau_0$ large \cite{cc-06,cc-07} 
\be 
\eta \sim \frac{e^{\pi t/\tau_0}}{e^{\pi \ell/2\tau_0}+ e^{\pi t/\tau_0}}\,. 
\ee 
For $t<\ell/2$, since $t,\ell\gg\t0$, we have $e^{\pi \ell/2\tau_0}\gg e^{\pi t/\tau_0}$ and so in the 
denominator above we can neglect $e^{\pi t/\tau_0}$ and $\eta$ vanishes as $\eta\simeq e^{\pi(t-\ell/2)/\t0}$.
Oppositely for $t>\ell/2$ we can neglect $e^{\pi \ell/2\tau_0}$ in the the denominator and 
we have $\eta\sim1$. 
Even if $ F(\eta)$ is generally unknown,
we conclude that we only need its behaviour close to $\eta\sim 0$ and $1$, that are
easily deduced from general scaling. Indeed when $\eta \sim1$ the two
points are deep in the bulk, meaning $F(1)=1$.
Instead for $\eta\ll 1$, from the short-distance expansion, we have
\be
F(\eta)\simeq (A^\Phi_b)^2\eta^{x_b},
\ee
where $x_b$ is the boundary scaling dimension of the leading
boundary operator to which $\Phi$ couples and $A^\Phi_b$ is the bulk-boundary
operator product expansion coefficient that equals the one introduced 
in Eq.~(\ref{onepoint}) (see \em e.g. \em Ref.~\cite{cl-91}).
Also the prefactor to $F(\eta)$ assumes two simple limiting values for $t>\ell/2$ and 
$t<\ell/2$ when $t,\ell\gg \t0$ which are proportional to $e^{-x\pi \ell/2\t0}$ and $e^{-x\pi t/\t0}$ respectively. 

All the previous observations lead to
\be\fl 
\langle\Phi(\ell,t)\Phi(0,t)\rangle  \simeq  \left(\frac{\pi}{2\tau_0}\right)^{2 x}\times
\cases{
 e^{-x\pi \ell/2\t0}\,,&  for $t>\ell/2$,\cr
 (A^\Phi_b)^2 e^{-x\pi t/\t0}    e^{\pi x_b(t-\ell/2)/\t0}\,, &  for $t<\ell/2$. 
 }
\label{twocft}
\ee
Note that if $\langle\Phi\rangle\neq0$, $x_b=0$ and the last factor is absent. 
The leading term is then just $\langle\Phi\rangle^2$. 
Thus the leading term in the connected two-point function vanishes for $t<\ell/2$.

\subsection{Evolution of entanglement entropy after a global quantum quench} 
\label{entaglob}

The entanglement entropy of a finite interval $A$ of length $\ell$ in an infinite system 
can be easily obtained from the previous results by using the replica trick. 
Indeed, $\Tr \rho_A^n$ is equivalent to the two-point function of twist operators 
(which under conformal transformations behaves like primary operators) of dimension  \cite{cc-04,cc-09}
\be
x_n=\frac{c}{12}\Big(n-\frac1n\Big),
\label{twistx}
\ee 
where $c$ is again the central charge.
Thus from Eq.~(\ref{twott}) we have,  in the case where $\ell/\tau_0$ and $t/\tau_0$ are
large, the moments of the reduced density matrix simplify to
\begin{equation}
{\rm Tr}\,\rho_A^n(t)\simeq
c_n(\pi/2\tau_0)^{2x_n}\left(\frac{e^{\pi\ell/2\tau_0} +e^{\pi
t/\tau_0}} {e^{\pi\ell/2\tau_0}\cdot e^{\pi t/\tau_0}}\right)^{x_n}\,,
\label{momgc}
\end{equation}
where we explicitly use that the function of the four-point ratio $F(\eta)$ is equal to 1 for both 
$\eta\sim 0$ and $\eta\sim1$ \cite{cc-09} which are the only two cases of interest. 
The constants $c_n$ are non-universal normalisations \cite{cc-04}.

At this point, the entanglement entropy $S_A=-\Tr \rho_A \ln \rho_A$
can be obtained by continuing the above result to arbitrary real values of $n$,
differentiating wrt $n$, and taking $n\to1$. 
After simple algebra, we obtain \cite{cc-05} 
\be S_A(t)\simeq
\frac{c}3 \ln \tau_0+ 
\cases{ \displaystyle      \frac{\pi c t}{6\tau_0}    \qquad t<\ell/2 \;,
\cr \displaystyle      \frac{\pi c\,\ell}{12\tau_0}\qquad t>\ell/2\,, } \label{SAt2} 
\ee 
that is $S_A(t)$ increases linearly until it saturates at $t=\ell/2$. 
The sharp cusp in this asymptotic result is rounded over a region $|t-\ell/2|\sim\tau_0$. In contrast with Ref.~\cite{cc-05},
following \cite{sc-08}, we have added explicitly the subleading
constant term $\ln\tau_0$ confirming that $\tau_0$ is proportional
to the correlation length in the initial state, given that in an equilibrium gapped system at zero temperature  the 
entanglement entropy is $c/3 \ln \xi$ \cite{cc-04}.

From Eq.~(\ref{momgc}) it follows that the time dependence of all the R\`enyi entropies
$S_A^{(n)}\equiv (\ln \Tr\rho_A^n) /(1-n)$ is of the form (\ref{SAt2}), but multiplied by a $n$-dependent constant.

\subsubsection{General result for an arbitrary number of intervals.}
A general result can be also derived in the case when
$A$ consists of the union of the $N$ intervals $(u_{2j-1},u_{2j})$ where $1\leq j\leq N$ and
$u_k<u_{k+1}$. ${\rm Tr}\,\rho_A^n$ is given by a $2N$-point function of twist operator in a strip of width $2\tau_0$. 
We only need the asymptotic behaviour of this correlation function for time $t$ and
separations $|u_j-u_k|$ much larger than $\tau_0$. Several simplifications happen in this regime and after 
long algebra one arrives at \cite{cc-05}
\begin{equation}
S_A(t)\sim S_A(\infty) +\frac{\pi c}{12\tau_0}
\sum_{k,l}(-1)^{k-l-1}{\rm max}(u_k-t,u_l+t)\,.
\label{Smany}
\end{equation}
If $N$ is finite (or more generally the $u_k$ are bounded) the
second term vanishes for sufficiently large $t$. At shorter times,
$S_A(t)$ exhibits piecewise linear behaviour in $t$ with cusps
whenever $2t=u_k-u_l$, at which the slope changes by $\pm\pi c/6\tau_0$
according to whether $k-l$ is even or odd. In the
case of an infinite number of regular intervals, with $u_k=k\ell$,
$k\in{\rm Z}$, $S_A(t)$ exhibits a sawtooth behaviour.
Eq.~(\ref{Smany}) has been checked numerically against exact results in the critical harmonic chain \cite{ctc-15} 
finding excellent agreement. However, it has recently been pointed out \cite{abgh-15,sm-15} that for a CFT with an infinite number of primary fields (\em e.g. \em CFTs with $c>1$ and only Virasoro symmetry) the valleys in $S_A(t)$ are partly filled in, and indeed they do not exist at all in holographic theories with $c\gg1$. This may be traced to the existence of additional singularities in the twist field correlators in the real time domain, beyond those implied by the usual OPE.

\subsection{Thermalisation when starting from $e^{-\tau_0 H} |\psi_0^* \rangle$}
\label{therm}

The results presented in the previous subsections for the one- and two-point functions of primaries 
are such that for large times they are those at finite temperature $T$ with $\beta=1/T= 4\tau_0$.
Also the extensive part of the entanglement entropy of a subsystem with its complement 
equals for large time the Gibbs entropy at the same temperature. 
We have also shown in Eq.~(\ref{Ttt}) that the local energy density (being non primary) does not relax. 
But, we also know that at finite temperature $\langle T_{tt}\rangle_\beta= \pi c/6 \beta^2$ \cite{BCN, Affleck}, 
that is perfectly compatible with a long time thermal expectation with $\beta=4\tau_0$.
It has been shown in \cite{c-15} that arbitrary multipoint correlation of primaries are indeed described by the thermal 
ensemble. 
Overall these results suggest that the system should be locally described by a thermal ensemble at a temperature 
corresponding to the conserved energy density. 

In \cite{cc-07}, we pointed out a simple \em technical \em reason why we find an effective temperature for long time.
The finite temperature correlations can be calculated by studying the field theory on a cylinder of circumference $\beta$. 
In CFT a cylinder can be  obtained by mapping the complex plane with the logarithmic transformation $\beta/(2\pi)\ln z$.
Focusing for simplicity of the two-point function of primaries, their form in the strip depends in general on the 
function $F(\eta)$ -- \em cf. \em Eq.~(\ref{2ptgen}) -- but when we analytically continue  and  take the limit of large real time, we find that 
effectively the points are far from the boundary, i.e. at $\eta\approx1$. 
Thus we get the same result as we would get if we conformally transformed 
from the full plane to a cylinder, and from Eq.~(\ref{logmap}) the effective temperature is $\beta=4 \tau_0$.
A similar argument can be worked out for the multi-point functions as well, see \cite{c-15} for details.  

However, the above argument leaves open the question of other correlation functions, such as those 
of non-primary operators. 
To fill this gap it was pointed out in \cite{c-15} that the reduced density matrix of an interval is close to that of a 
thermal ensemble, once the interval has fallen inside the horizon. 
This then implies that all equal-time correlation functions of local operators (with arguments in the interval) 
are close to their thermal values. 

The reasoning of Ref.~\cite{c-15} proceeds as follows. 
Consider an interval $A$ of length $\ell\ll L$ (in practice it sufficec that $L/2-\ell\gg\beta$). 
The reduced density matrix $\rho_A$ is 
\be
\rho_A=\frac{{\rm Tr}_{\overline A} [e^{-H(\tau_0+\tau)}|\psi_0^*\rangle\langle \psi_0^*|e^{-H(\tau_0-\tau)}]}{{\rm Tr}\,e^{-\tau_0H}|\psi_0^*\rangle\langle \psi_0^*|e^{-\tau_0H}}\,,
\ee
continued to $\tau=it$. 
For real $\tau$ the numerator of this expression is the partition function on a strip ${\cal S}$ slit along $(\ell,\tau)$.
The denominator is the trace over $A$ of this, which is equivalent to sewing up the slit. 
Similarly the reduced density matrix in a thermal ensemble 
\be
\tilde\rho^\beta_A=\frac{{\rm Tr}_{\overline A}\,e^{-\beta H}}{{\rm Tr}\,e^{-\beta H}},
\ee
is the partition function on a cylinder $\cal C$ slit along $(\ell,\tau)$, divided by the partition function on the full cylinder. 

The closeness of these two reduced density matrices is given by the overlap
\be
\frac{{\rm Tr} (\rho_A\cdot\tilde\rho^\beta_A)}{\sqrt{({\rm Tr} \rho_A^2)({\rm Tr}\,(\tilde\rho_A^\beta)^2)}}
=\frac{Z({\cal S}\oplus{\cal C})}{\sqrt{Z({\cal S}\oplus{\cal S})Z({\cal C}\oplus{\cal C})}}\,,
\label{ratio}
\ee
where $Z({\cal S}\oplus{\cal C})$ is the partition function on $\cal S$ sewn onto $\cal C$, 
in such a way that the bottom edge of the slit in $\cal S$ is sewn to the top edge of $\cal C$ and vice versa. 
As already mentioned, in a CFT  these partition functions may be viewed as correlators of twist operators  
$\cal T$ evaluated in the product of the CFTs on each component \cite{cc-04,twist}, i.e. 
\be
\fl\frac{Z({\cal S}\oplus{\cal C})}{\sqrt{Z({\cal S}\oplus{\cal S})Z({\cal C}\oplus{\cal C})}}
=\frac{\langle {\cal T}(x_1,\tau){\cal T}(x_2,\tau)\rangle_{{\cal S}\otimes {\cal C}}}
{\sqrt{\langle{\cal T}(x_1,\tau){\cal T}(x_2,\tau)\rangle_{{\cal S}\otimes {\cal S}}\langle{\cal T}(x_1,\tau){\cal T}(x_2,\tau)\rangle_{{\cal C}\otimes {\cal C}}}}\,,
\label{ratio2}
\ee
where now $\langle {\cal T}(x_1,\tau){\cal T}(x_2,\tau)\rangle_{{\cal S}\otimes {\cal C}}$ means the twist correlator on 
a direct product of the CFT on $\cal S$ with that on $\cal C$.

The twist operators enjoy the same analyticity properties of primary operators in a CFT and  
therefore we may use the earlier arguments applied to correlators of primaries to compute (\ref{ratio2}). 
Applying the conformal mapping (\ref{logmap}), they are related to correlators of twist operators on CFTs on 
${\mathbb H}\otimes{\mathbb C}, {\mathbb H}\otimes{\mathbb H},{\mathbb C}\otimes{\mathbb C}$, 
with ${\mathbb H}$ being the upper-half plane.
Once the points $(x_1,\tau)$ and $(x_2,\tau)$ have fallen into the horizon, the cross-ratio $\eta_{12}$ is $\ll1$, 
and we may use the operator product expansion (OPE).

The OPE of twist operators is expressed as a sum of products of local 
operators in each sheet as \cite{CCT,Head}
\be
{\cal T}(z_1)\cdot{\cal T}(z_2)=\sum_{k_1,k_2}C_{k_1,k_2}\Phi_{k_1}\Phi_{k_2}\,,
\ee
where $\Phi_{k_1}$ and $\Phi_{k_2}$ are a complete set of operators (arranged in order of increasing dimension) 
in copies of the CFT. 
The coefficients $C_{k_1,k_2}$ are calculable, but their values are not interesting for the present calculation. 
The leading term in (\ref{ratio2}) comes from taking $\Phi_{k_1}=\Phi_{k_2}={\bf 1}$, so that, once the ends of the interval 
have fallen inside the horizon, Eq.~(\ref{ratio2}) is asymptotically equal to one. 
The leading correction comes from the operators $(\Phi_{k_1},\Phi_{k_2})$ with the 
lowest dimension and such that they have non-zero expectation values. 
The lowest dimension operators are usually primary which have vanishing expectation value in $\mathbb C$. 
Consequently, the leading corrections come only from the first factor in the denominator of (\ref{ratio2}) and correspond to the most 
relevant operators which have a non-vanishing expectation values on $(\mathbb H,\mathbb H)$ 
given the particular boundary state $|\psi_0^*\rangle$. 
Denoting with $\Delta$ the dimension of this operator, we have
\begin{equation}\label{close}
1-\frac{{\rm Tr}\,(\rho_A\cdot\tilde\rho^\beta_A)}{\big(({\rm Tr}\,\rho_A^2)({\rm Tr}\,(\tilde\rho_A^\beta)^2)\big)^{1/2}}
<{\rm const.}\,e^{-2\pi\Delta(t-\ell/2v)/\beta}\,,
\end{equation}
showing that long time after the quench the reduced density matrix of an arbitrary interval is 
exponentially close to the reduced density matrix of a thermal ensemble. 

The thermalisation found in this section is however rather strange since conformal field theories are prototypes of quantum 
integrable models. Due to the many local or quasi-local conservations, it is indeed generically expected that for 
large times the system should converge to a generalised Gibbs ensemble (GGE) in which all the integral of motions are 
present. 
As firstly pointed out in \cite{c-15}, this is just a due to the very special initial state (\ref{tau0B}) we considered so far.
 Indeed, by perturbing this initial state with irrelevant boundary operators, it can be shown that the asymptotic state 
is a GGE rather than a thermal one. We will report explicitly this argument in Sec.~\ref{genIS}.

\subsection{Light-cone spreading of entanglement and correlations}
\label{lc-sec}

\begin{SCfigure}%[t]
\includegraphics[width=0.5\textwidth]{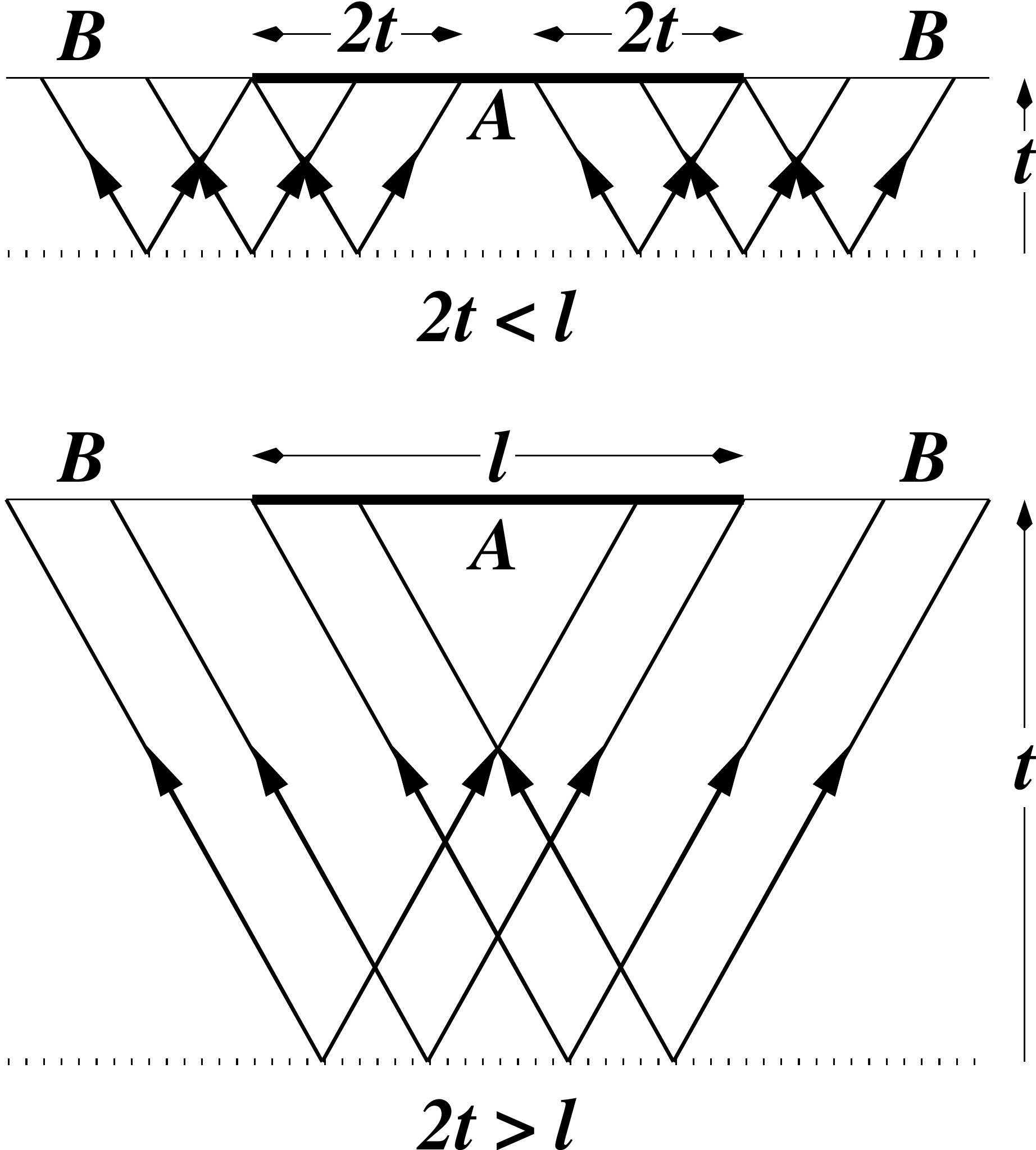}
\caption{
The linear growth of the entanglement entropy followed by saturation can be 
understood in terms of oppositely moving coherent quasiparticles as pictorially reported here. 
Reprinted with permission from \cite{cc-05}.}
\label{fig2tl}
\end{SCfigure}

We have seen in the previous section that the correlation functions and the entanglement 
display a non-analytic behaviour on the light-cone at $t=\ell/2$ for a conformal post-quench evolution.
These non-analyticities can be understood as a {\it light-cone like spreading} of entanglement and correlation which 
can be physically motivated by the following argument (firstly proposed in \cite{cc-05} and later generalised in \cite{cc-06,cc-07}).
This scenario is generically valid for a large class of quenches starting from a state with short-range correlations and 
entanglement (such as the ground state of a gapped hamiltonian) and it is not restricted to conformal invariant 
post-quench hamiltonians.
We firstly explain how the argument works for the entanglement entropy and after for correlations. 

A generic initial state $|\psi_0\rangle$ for a global quantum quench has a very high energy relative to
the ground state of the post-quench hamiltonian, and therefore acts as a source of quasiparticle excitations.
Particles emitted from different points (further apart than the
correlation length in the initial state) are incoherent,
but pairs of particles moving to the left or right from a given point are  entangled. 
We denote with $\rho(p)$ the rate of production of such a pair of particles of momenta $(p,-p)$, 
The function $\rho(p)$  depends on the post-quench hamiltonian $H$ and on the initial state $|\psi_0\rangle$, and in
principle is calculable, but we do not make any assumptions on its form.
We instead assume that, once the two quasiparticles separate, they move classically.
If the quasiparticle dispersion relation is $E=E_p$, the  velocity is $v_p=dE_p/dp$.
A quasiparticle of momentum $p$ produced at $x$ is therefore at $x\pm v_pt$ at time $t$, ignoring scattering effects.
We assume that there is a maximum allowed speed $v_{\rm max}$, i.e. $|v_p|\leq v_{\rm max}$. 
There are various reasons to assume so: in a lattice model this is a consequence of Lieb-Robinson bound \cite{lr-72},
in a relativistic field theory it follows from causality, and in some other cases it could 
just happen that the quench populates only modes with a finite velocities. 

Let us now bipartition the one-dimensional system in two complementary parts $A$ and $B$. We are interested in the 
entanglement entropy between them. We then consider these quasiparticles as they reach either $A$ or $B$ at time $t$. 
The field at some point $x'\in A$ will be entangled with that
at a point $x''\in B$ if a pair of entangled particles emitted from a
point $x$ arrive simultaneously at $x'$ and $x''$ (see Fig.~\ref{fig2tl} for the case of $A$ consisting of a single interval).
We assume that the entanglement entropy between $x'$ and $x''$
is proportional to the length of the interval in $x$ for which this can
be satisfied. Thus the total entanglement entropy is
\be\fl
S_A(t)\approx \int_{x'\in A} 
dx'\int_{x''\in B} \hspace{-3mm}  dx''\int_{-\infty}^\infty
dx\int
f(p)dp\delta\big(x'-x-v_{p}t\big)\delta\big(x''-x+v_{p}t\big),
\ee
where $f(p)$ is equal to the product of $\rho(p)$ with the contribution of a given pair to the entanglement entropy. 
Now we specialise to the case where $A$ is an interval of length $\ell$.
The integrations over the coordinates are easily done, so that
\bea
 S_A(t)&\propto& t \int_{2 v_p t <\ell } dp f(p) 2 v_p  +\ell \int_{2 v_p t >\ell } dp f(p)\,. \label{ppp}
\eea
Since $|v_p|\leq v_{\rm max}$, the second term cannot contribute if $t<t^*=\ell/2v_{\rm max}$, so that $S_A(t)$ is
strictly proportional to $t$. On the other hand as $t\to\infty$, the
first term is negligible (this assumes that $v_p$ does not vanish
except at isolated points), and $S_A$ is asymptotically proportional to $\ell$.
However, unless $|v|=v_{\rm max}$ everywhere (as is the case of a CFT with $v_{\rm max}=1$),
$S_A$ is not strictly proportional to $\ell$ for $t>t^*$, since the first term in (\ref{ppp}) does not vanish. 
The approach to the stationary value depends on the behaviour of $f(p)$ in the regions where $v_{p}\to 0$. 
This generally happens at the zone boundary, and, for a non-critical quench, also at $p=0$. 
Since usually close to its zeroes $p_i$, $v_p$ is linear $v_p\sim (p-p_i)$, and assuming $f(p)$ to be regular at $p_i$,
we expect that for large times the first term in (\ref{ppp}) vanishes as $\ell^2/t$. 
However, this can change depending on the explicit form of $v_p$ and $f(p)$. 

We mention that also the complicated form (\ref{Smany}) for the entanglement entropy of $N$ disjoint intervals 
can be easily understood in terms of this picture with $v_p=1$ for all $p$ and can be generalised to 
quasiparticles with different velocities \cite{cc-05}. 
Furthermore also the R\`enyi entropy for arbitrary $n$ have the same form (\ref{ppp}) but with a 
$n$-dependent function $f_n(p)$.
It is also possible to quantify in specific lattice models the contribution to the entanglement entropy of two given points
in $A$ and $B$ by means of the so-called entanglement contour \cite{cv-14}. 
Also its time dependence after a quench is compatible with this quasi-particle picture in the case of free fermionic 
models, a case for which it has been explicitly checked \cite{cv-14}.
Some results for the spreading of entanglement are also available in higher dimensions \cite{clm-16}.

The same picture  explains also the behaviour of correlation functions of local operators. 
Indeed, entangled quasi-particles arriving at the same time $t$ at points with separation $\ell$ 
induce also correlations between local observables. 
In the case when they travel at a unique speed $v$ (as in CFT), there is a sharp light cone and connected correlations 
do not change significantly from their initial values until time $t\sim  \ell/2v$. 
In the CFT case this light-cone effect is rounded off  over the region $t- \ell /2v\sim\tau_0$,
since quasi-particles remain correlated over this distance scale.
After this they  saturate to time-independent values. 

In any realistic condensed matter model the precise time dependence of the correlators depends on many details of the 
theory and of the considered operators. In particular there is no degree of universality since the quench populates high 
energy modes. However, the above quasi-particle picture can be used as a starting point for a semiclassical approach to quenches 
as for example done in \cite{bri-12, ri-11,ir-11,e-13,kz-16}. This semiclassical approach allow to calculate 
time-dependent correlation functions in the limit of low density of excitations after the quench, which in some cases 
turned out to reproduce quite accurately also the finite density correlations, see \em e.g. \em  \cite{ir-11} for some comparisons 
of the semiclassical against exact computations.

\subsection{Some further results in 1+1 conformal global quenches}

In this subsection we briefly mention a few other results for global quantum quenches that can be derived 
with the imaginary time formalism of the previous subsections, but that for lack of space we cannot review properly. 

In Ref. \cite{cc-06,cc-07}, the two-point function of primaries at different times was explicitly computed, obtaining  
(in the case with $F(\eta)=1$)
\be
\langle\Phi(r,t)\Phi(0,s)\rangle=
\cases{
e^{-x\pi(t+s)/4\t0}\quad &{\rm for}\, $r>t+s$\,,\cr
e^{-x\pi r /4\t0}\quad &{\rm for}\, $t-s<r<t+s$\,,\cr
e^{-x\pi|t-s|/4\t0}\quad &{\rm for}\, $r<|t-s|$\,.
}
\ee
These results were generalised to the two-time response function in \cite{mg-14}.

Quantum quenches in the presence of boundaries were also studied in the framework of boundary CFT \cite{cc-07,dmcf-06}, 
but the results are very similar to those in the bulk and we do not discuss them here. 

In \cite{ctc-15}, the time evolution of the entanglement entropy, of the mutual information 
and of the so called entanglement negativity of many disjoint intervals has been investigated in great details.
The obtained results are compatible with the quasi-particle picture for the spreading of entanglement, 
see also \cite{nr-14}.

The imaginary time formalism can be used also to treat massive field theories, but the results in this case 
are much less general as, for example, discussed  in \cite{fm-10}.

\subsection{Insights and differences when comparing with realistic condensed matter models}

Contrary to naive expectations, the results derived in the previous sections for conformal field theories 
do not always apply to critical models whose low-energy physics is described by a CFT, as it is the case in equilibrium at low enough temperature. 
In the literature the CFT predictions for global quenches have been often taken 
far beyond their regime of applicability, especially when comparing with results for lattice models.  
One of the main drawbacks that is often overlooked is that CFT gives  an accurate description only of the 
low energy properties of (some) gapless one dimensional quantum models. 
However, in performing a global quench, a large (indeed extensive) amount of energy is injected  into the system 
and generically one populates states in the middle of the many-body spectrum, a region which is well beyond 
the regime of applicability of CFT. 
Another drawback is that in a CFT all the quasi-particle excitations move with the same speed (which here has been fixed to unity)
independently of their energy. 
As already stressed in the subsection about light-cone spreading \ref{lc-sec}, this is not the case for a generic gapless model, 
even with a free particle description (such as the tight binding model or the critical Ising model). 
Indeed, while for small momentum $k$, the dispersion relation $E_k$ has a CFT form $E_k\sim v |k|$, 
for larger values of $k$ it becomes a non-trivial function of the momentum.

The last problematic issue concerns the initial state we are considering, in particular
the identification of $\tau_0$ which should be handled carefully. 
Indeed, we have argued (see also \cite{cc-06,cc-07}) that, when the initial field-theoretical correlation length $\xi_0$ is small, we 
roughly expect $\tau_0\propto \xi_0$. 
However, when comparing with a lattice model, the field theoretic description is appropriate only when $\xi_0$ is much 
larger that the lattice spacing. Mathematically, one should first consider the scaling limit in which the correlation 
length in terms of the lattice spacing diverges and later consider distances and times much larger than the resulting 
field-theoretical correlation length. Physically (and in practice), one should hope that for a given lattice model 
there is a regime in which $a\ll\xi_0\ll \ell,t$.
Thus even in models which are described to a great level of accuracy by a CFT (such as a lattice Dirac fermion 
or the harmonic chain in specific regimes), the initial state plays a crucial rule for finding results compatible with 
the picture highlighted in the previous subsections. 
A more general class of initial states will be explicitly considered in Sec.~\ref{genIS}.

\begin{figure}%[t]
\includegraphics[width=0.45\textwidth]{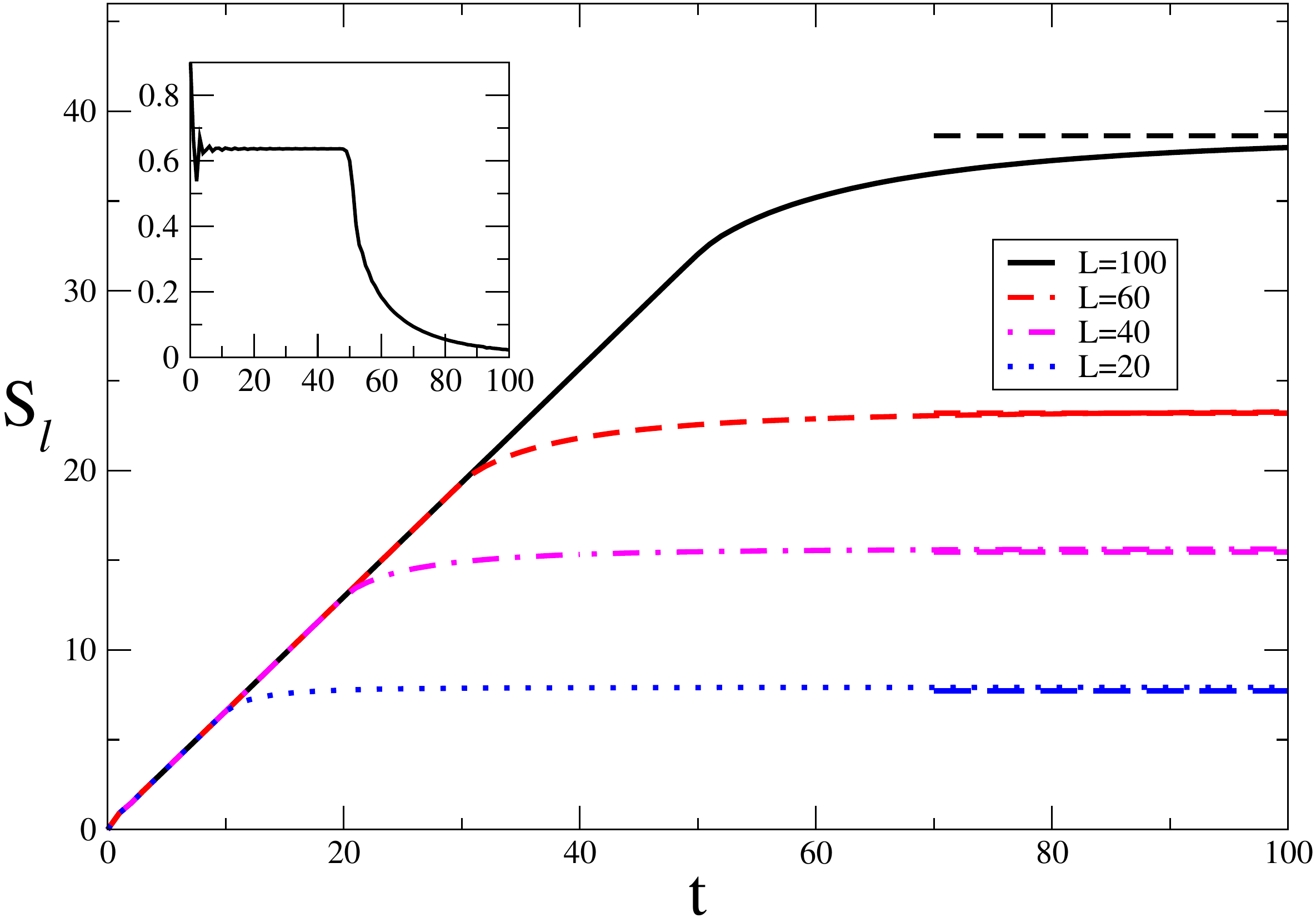}
\includegraphics[width=0.45\textwidth]{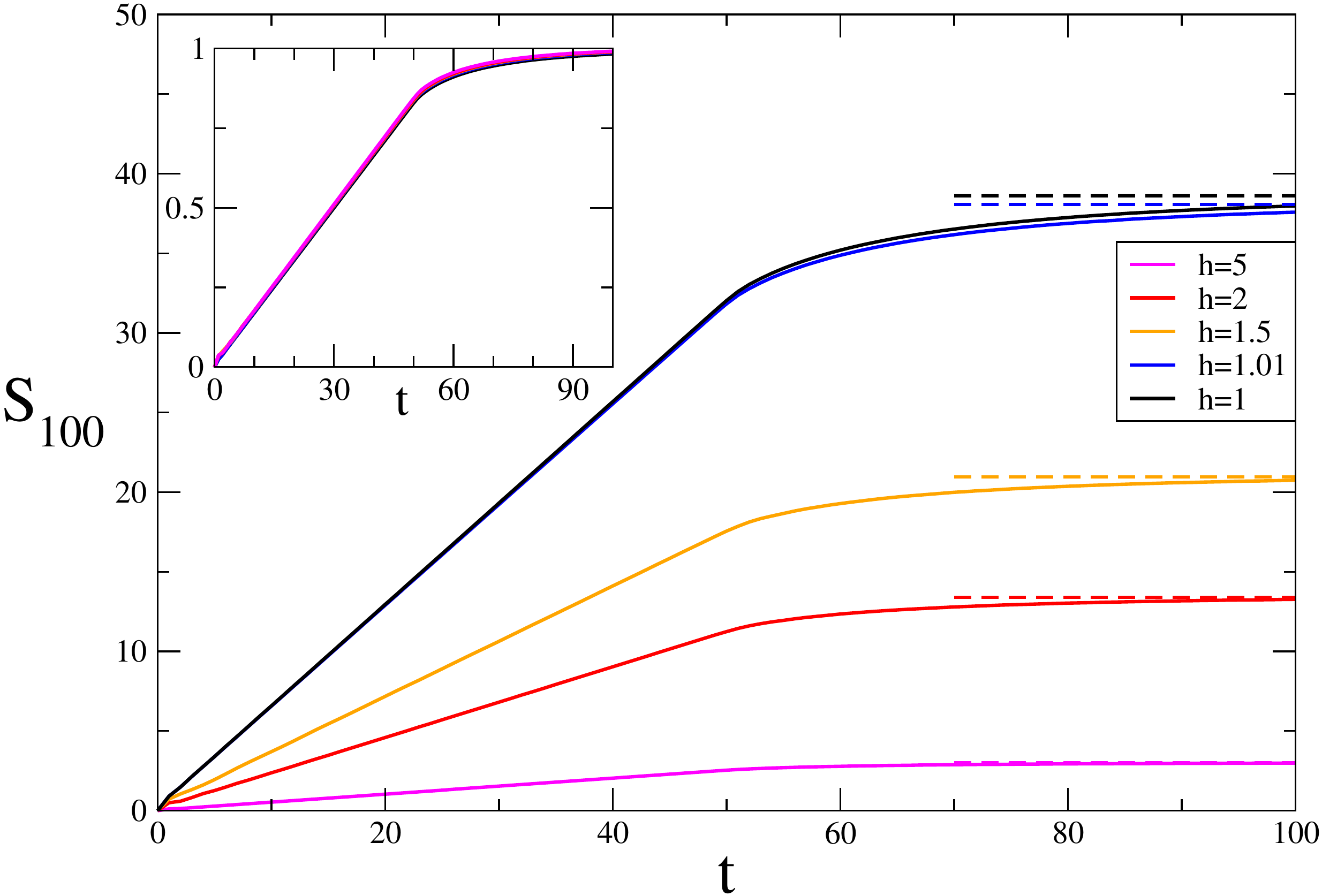}
\caption{Exact entanglement entropy in the the transverse field Ising model in the thermodynamic limit. 
Left: Time evolution of the entanglement entropy as function of time for a quench from $h_0=\infty$ to the 
critical point $h=1$ for different lengths $L$ of the subsystem.  
Right: The same but for fixed length of the subsystem ($L=100$) and as a function of the post-quench magnetic field $h$.
Both panels reprinted with permission from \cite{cc-05}.} \label{EEising}
\end{figure}

After having listed the main drawbacks in trying to quantitatively compare the CFT predictions for global quenches with explicit results 
in condensed matter 
systems, we can proceed to show some of these results and discuss what we can learn from CFT for more realistic models. 
We start by showing the results for the time evolution of the entanglement entropy in a quench in the transverse field Ising chain 
described by the hamiltonian
\be
H(h)=-\frac12 \sum_{l=-\infty}^\infty \left[
 \sigma_l^x\sigma_{l+1}^x+h \sigma_l^z \right]\,,
\label{Hamil}
\ee
which at zero temperature exhibits ferromagnetic ($h<1$) and
paramagnetic ($h>1$) phases, separated by a quantum critical point described by a CFT (see \cite{sach-book} 
for the description of the equilibrium behaviour of this model).
We report results for the quench from a given transverse field $h_0$ to another value $h\neq h_0$.
In the left panel of Fig.~\ref{EEising} we show the results for a quench to the critical point $h=1$ starting from 
the product state at $h_0=\infty$ (to mimic a boundary conformally invariant state). 
We can see that the entanglement entropy grows  linearly for $t<\ell/2$, in agreement with CFT and the 
quasiparticle picture with the known sound velocity $v_{\rm max}=1$. 
For $t>\ell/2$, the entanglement entropy does not saturate abruptly as in a CFT, but continues to increase slowly. 
We can  understand this as the effect of the slower quasi-particles as explained in Sec.~\ref{lc-sec}. 
In the right panel of Fig.~\ref{EEising} we report the time evolution of the entanglement entropy for an interval of fixed length, 
starting always 
from the initial state with $h_0=\infty$ and the various curves correspond to different values of the final value of the field $h$.
It is evident that all curves show the same behaviour, characterised by the same velocity $v_{\rm max}=1$, but with a
rate of growth of the entropy  depending on the quench parameters. 
Again this behaviour is fully compatible with the quasi-particle picture and it shows how the linear increase of the 
entanglement entropy, followed by an almost saturation is a generic feature of many quenches and it is not restricted 
to conformal invariant and gapless situations. 
For a general quench in the Ising model, indeed the time dependence of the entanglement entropy in the 
so called space-time scaling regime (i.e. for $t, \ell\to\infty$ with the $t/\ell$ kept fixed) has been derived exactly \cite{fc-08}, 
obtaining 
\be
S_A=t \int\limits_{2 v_k t<\ell}\frac{d k}{2\pi} 2 v_k H(\cos\Delta_k) +\
\ell \int\limits_{2 v_k t>\ell}\frac{dk }{2\pi} H(\cos\Delta_k)\,,
\label{St}
\ee
where $v_k= h \sin k/\sqrt{1+h^2-2h \cos k}$ is the velocity of the $k$-mode,
$\cos \Delta_k$ is a known function of $k$ which encodes  all quench information (see for its definition \cite{fc-08})
and  $H(x)=-((1+x)/2\ln(1+x)/2+(1-x)/2\ln(1-x)/2)$. 
This compact analytic formula shows that the quasi-particle picture not only gives a qualitative description of the 
dynamics, but also provides accurate quantitative prediction such as (\ref{ppp}) which coincides with (\ref{St}) 
after identifying $f(p)$ with $H(\cos\Delta_k)$.
 
Indeed, the linear growth of the entanglement entropy followed by an almost saturation has been observed numerically 
in  a very large number of exact calculations and in quantum simulations, such as in 
\cite{dmcf-06,nr-14,fc-08, lk-08,ep-08,kh-13,ckc-14,bkc-14,fc-15,pbr-16} 
(but this list is far from being exhaustive). 
The standard quasi-particle picture breaks down in models with disorder, for which it has been shown that 
the growth of entanglement is logarithmic in time (or slower) \cite{dmcf-06,bo-07,isl-12,bpm-12,va-14,zas-16}, 
and in model with long-range interactions \cite{ht-13,slrd-13, jur-14,ric-14,bf-16}. 

The linear growth of the entanglement entropy in time is also a very important physical phenomenon to understand why 
algorithms based on matrix product states \cite{mps} (such as the  density matrix renormalisation 
group \cite{dmrg} in its time dependent version \cite{tDMRG} or iTEBD \cite{iTEBD}) 
fail to describe the quench dynamics for  large times. 
Indeed, it has been established that a matrix product state built with a tensor of auxiliary dimension $\chi$ can 
effectively encode a quantum state whose entanglement entropy is of order of $\log \chi$ 
(see \em e.g. \em Refs.~\cite{swvc-08,chisca} for a more precise statement). 
Given that the entanglement entropy after a global quench grows linearly in time, one would need that the auxiliary dimension 
of the matrix product state should grow exponentially with time  and this contrasts with the limited computational 
resources at disposal.

\begin{figure}[t]
\includegraphics[width=0.48\textwidth]{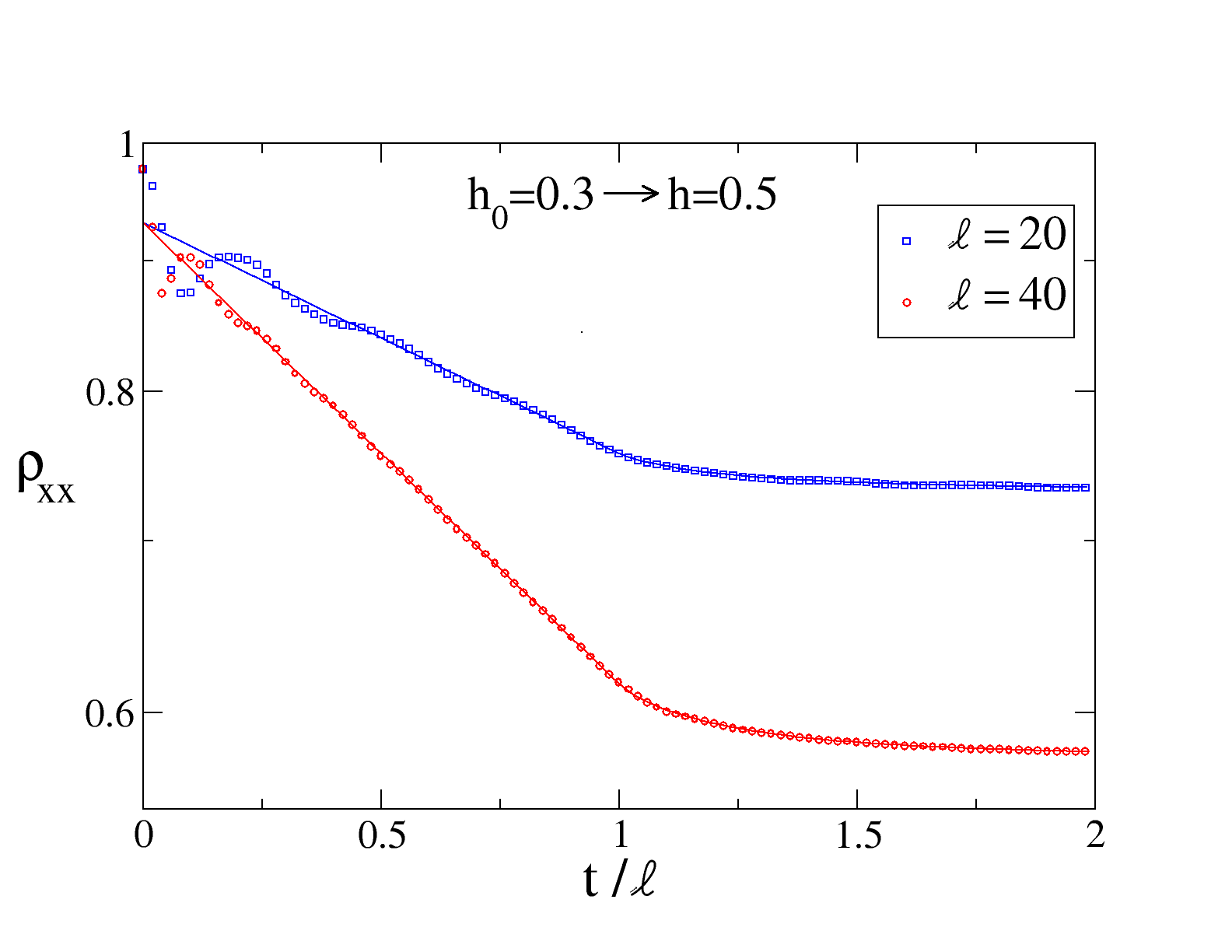}
\includegraphics[width=0.48\textwidth]{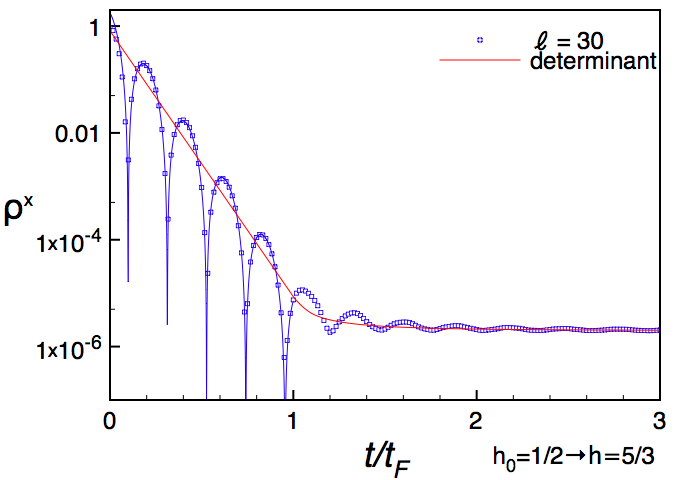}
\caption{Equal time two-point correlation function following a quench in the Ising chain. 
Left: a ferro-to-ferro quench $h_0=0.3\rightarrow h=0.5$ at fixed distance $\ell=20$ and $\ell=40$ against the  prediction 
(\ref{eq:prediction}). 
Right: Ferro-to-para quench showing large oscillations inside the light cone.
Reprinted with permission from \cite{cef-11} and \cite{cef-12a}. 
} 
\label{fig:1}
\end{figure}

Let us now discuss the behaviour of correlation functions and let us again start by reviewing some exact results 
for the Ising chain (\ref{Hamil}), whose correlation functions after a quench have been determined analytically 
in \cite{cef-11,cef-12a,cef-12b,eef-12,fe-13,se-12} (see \cite{mc,ir-00,sps-04,rsms-08,fcg-11} for earlier numerical works).
When starting from the ferromagnetic phase (i.e. $|h_0|<1$) 
in such a way to have a non-zero initial order parameter, and quenching to a point with $h\leq 1$,
 it has been shown that at late times the order parameter relaxes to zero exponentially fast as \cite{cef-11,cef-12a}
\begin{equation}
\label{eq:OP}
\langle \psi(t)|\sigma^x_j|\psi(t)\rangle  \propto\exp\left[t \int_{0}^\pi \frac{\mathrm d k}{\pi} v_k \ln\left(\cos\Delta_k\right) \right],
\end{equation}
a behaviour which is reminiscent of the CFT prediction (\ref{onepoint}).
This exponential decay has been indeed found also in other, even interacting, models \cite{bpgda-10,bse-15}.
For the same quench, the equal time two-point function of the order parameter in the space-time scaling limit
 has been found to be  \cite{cef-11,cef-12a}
\be\fl
\label{eq:prediction}
\langle \psi(t)| \sigma^x_\ell \sigma^x_1 |\psi(t)\rangle \propto\exp
\Bigg[\ell \int\limits_{2 v_k t>\ell} \frac{\mathrm d k}{\pi}\ln\left(\cos\Delta_k\right)
+%\Big] \exp\Bigg[
2t \int\limits_{2 v_k t<\ell}\frac{\mathrm d k}{\pi} v_k
\ln\left(\cos\Delta_k\right)\Bigg].
\ee
This exact results shows several important properties of the quench dynamics. 
Since $v_k$ is bounded $|v_k|<v_{\rm max}$, the two-point function decays exponentially in time 
for $v_{\rm max} t<\ell$ and then saturates slowly to a value which is exponentially small in the separation. 
This is the same as the CFT behaviour (\ref{twocft}) but it is generically valid for any quench within the ferromagnetic phase 
up to  the critical point in the post-quench hamiltonian. 
In the limit $\ell\to\infty$ the first term vanishes showing that the two-point function is the square of the one-point one and so 
cluster decomposition holds during the quench dynamics. 
Furthermore, the first term is zero for all times such that 
$2 v_{\rm max} t<\ell$ and hence the connected correlation is vanishing in this regime.
This is a feature which was first shown in CFT, but which is generically valid and it goes nowadays under the name 
of {\it light-cone spreading of correlations} or alternatively {\it horizon effect}.
In Fig.~\ref{fig:1} numerical data are reported for the two-point function for quenches from the 
ferromagnetic phase to both the same phase (left) and to the paramagnetic phase (right).
In the former case, it is clear that (a part a short transient) the data are well described by  
Eq.~(\ref{eq:prediction}). In the latter instead it is evident that something new takes place since 
the relaxation of the two-point function is oscillating. 
This  has no analogy in the CFT calculations, but it is not unexpected since the post-quench hamiltonian is massive. 
We do not report plots for quenches from the paramagnetic phase (they can be found in \cite{cef-12a}), but we simply mention 
that in this case the dynamics is very complicated compared to the CFT one. 
The main important point to stress is that the connected correlation 
functions always vanish for $2 v_{\rm max} t<\ell$ (more precisely they are exponentially suppressed) and the 
light-cone spreading is valid.

As we already said, the light-cone spreading of correlations is a very general effect that was firstly understood 
in CFT \cite{cc-06,cc-07}.
Since then it has been observed in dozens of numerical simulations and exact calculations 
(see \em e.g. \em\cite{lk-08,mwnm-09,bpck-12,es-12,cbsf-14,bel-14,glms-14,cce-15})  
and finally it has been also experimentally demonstrated  in a cold-atomic setup \cite{cetal-12}.

The exact calculations for the Ising model show also one of the limits of the CFT calculations presented.
For large times, the two-point function (\ref{eq:prediction}) is not thermal. Indeed it has been shown 
that this correlator and the entire reduced density matrix of an arbitrary finite subsystem (and hence arbitrary correlations of 
all local operators)  are described by a generalised Gibbs ensemble (GGE) built with all the local charges of the 
model \cite{cef-12b,fe-13}.
In Sec.~\ref{genIS} we will see that this peculiar CFT behaviour is due to the particular choice of the initial state and that 
more general initial conditions will indeed approach for large times a GGE. 

Finally, it is worth mentioning that several CFT results for global quantum quenches both for the entanglement entropy and
correlations  have been re-obtained in the 
holographic framework of the AdS/CFT correspondence and generalised to higher dimensional 
CFTs \cite{BH,hrt-07,aal-10,aj-10,hol1,hol2,al-11,bbc-11,kkt-12,at-12,bmn-13,hrt-13,hm-13,ffk-14,ls-14,bmn-14,aam-14,hol-col,mss-15}. 
In this approach the thermalisation of a strongly coupled CFT is equivalent to the formation of a black hole in the 
AdS space, while a GGE is a higher-spin black hole \cite{mss-15}. 

\subsection{Revivals in finite systems}
\label{sec:rev}

We argued in Sec.~\ref{therm} that the density matrix of a interval of length $\ell$ 
become stationary after a time $\simeq\ell/2v$, after which it is described by a thermal ensemble at a temperature corresponding 
to the conserved energy density.
These considerations were made in the thermodynamic limit, but
the result can be extended \cite{c-14} to the case of finite total length of the system $L$ as long as 
$\ell/2<t<(L-\ell)/2$ (for periodic boundary conditions). 

However, according to the quasi-particle picture, in a conformal periodic (or open) system an oppositely moving 
pair of particles will meet again at times which are integer multiples 
of $L/2$ (respectively $L$), and this should generically lead to a revival of the initial state. 
In fact, in the transverse field XY spin chain  \cite{ir-11,ri-11,hhh-12} and in Luttinger liquids \cite{Caza} such revivals in the 
expectation values of local observables have been observed, and also in the entanglement 
entropy for a free Dirac fermion \cite{tu-10}. 

Here we describe the extension of the methods reviewed in previous sections to the case of finite systems 
as firstly presented in \cite{c-14}. 
We compute the return amplitude, or fidelity ${\cal F}(t)=|\langle\psi_0|e^{-iHt}|\psi_0\rangle|$, by relating it 
to the partition function of the CFT on an annulus (or rectangle for open boundary conditions) continued to complex 
values of its modulus or aspect ratio. 
These CFT partition functions are known in many cases and hence we are able to obtain several analytic results. 
We note in passing that in recent papers \cite{zeroes} a similar quantity has been studied 
for various lattice models, and its singularities interpreted as dynamical phase transitions at finite $t$. 
For the case of a CFT studied here, the singularities occur close to every rational value of $t/L$ and are simply related to 
full or partial revivals of the initial state.

As before, we take as initial state $|\psi_0\rangle\propto e^{-\tau_0 H}|\psi_0^*\rangle$, 
where $|\psi_0^*\rangle$ is a conformal boundary state, obtaining  the return amplitude 
\be
\!{\cal F}(t)=\!\left|\frac{\langle \psi_0^* |e^{-\tau_0 H}e^{-itH}e^{-\tau_0 H}| \psi_0^* \rangle}
{\langle \psi_0^*|e^{-\tau_0 H}e^{-\tau_0 H}|\psi_0^*\rangle}\right|
=\left|\frac{Z_{\cal A}(2\tau_0 +it,L)}{Z_{\cal A}(2\tau_0,L)}\right|,
\label{eq1}
\ee
where $Z_{\cal A}(W,L)$ is the partition function of the CFT on an annulus of width $W$ and circumference $L$, with conformal boundary conditions corresponding to $\psi_0^*$ on both edges. 
The form of $Z_{\cal A}$ for a CFT is \cite{JC89}
\be
Z_{\cal A}(W,L)=\sum_\Delta |B_\Delta|^2\chi_\Delta(q)=\sum_{\widetilde\Delta}n_{BB}^{\widetilde\Delta}
\chi_{\widetilde\Delta}(\tilde q)\,,
\label{eq2}
\ee
where $q\equiv e^{2\pi i\tau}=e^{-4\pi W/L}$, $\tilde q=e^{-2\pi i/\tau}=e^{-\pi L/W}$, and $\Delta,\widetilde\Delta$ 
label the highest weights of Virasoro representations which propagate across and around the annulus respectively. 
The functions $\chi_\Delta(q)=q^{-c/24+\Delta}\sum_{N=0}^\infty d_Nq^N$ are called the characters of the representations, 
and $d_N$ is their degeneracy at level $N$. 
The coefficients $B_\Delta$ are the overlaps between the boundary state $|\psi_0^*\rangle$ and the Ishibashi states \cite{Ishi}. 
The integers $n_{BB}^{\widetilde\Delta}$ are the number of states with highest weight $\widetilde\Delta$ 
allowed to propagate around the annulus with the given boundary conditions.
We assume $n_{BB}^0=1$. 
For minimal CFTs with $c<1$ there is a finite number of allowed values of scaling dimensions $\Delta$ and $\widetilde\Delta$ given 
by the Kac formula. 
For a more general rational CFT, the number of different values (mod $\bf Z$) is still finite, but for a general CFT 
with $c>1$ it is infinite and their mean density grows exponentially with $\Delta^{1/2}$ \cite{JC86}.  

The main needed property of the characters is that they are holomorphic in the upper half $\tau$-plane, 
and that they transform linearly under a representation of the modular group SL$(2,{\bf Z})$, 
generated by $S:\tau\to-1/\tau$ and  $T:\tau\to\tau+1$. 
The first property ensures that the continuation to $\tau=2(-t+i\tau_0)/L$ implied in (\ref{eq1}) makes sense, 
and the second  allows us to relate the values of ${\cal F}(t)$ at different times to those back in 
the principal domain where $\tau\to i\infty$ and the series are rapidly convergent.

Note that $\tilde q=\exp(-\pi L(\tau_0-it)/(\tau_0^2+t^2))$.
For $t^2\ll L\tau_0$, $|\tilde q|\ll1$, and so the sum on the rhs of (\ref{eq2}) is dominated by its first term ${\tilde q}^{-c/24}$. 
After normalising by the denominator in (\ref{eq1}) this gives
\begin{equation}\label{eq3}
{\cal F}(t)\simeq \exp (-(\pi c/24)Lt^2/\tau_0(\tau_0^2+t^2)) (1+O(|\tilde q|^{\alpha}))\,,
\end{equation}
which shows a decay, initially faster than exponential, to a plateau value which is however exponentially small in $L/\tau_0$. 
The power $\alpha$ in the correction term is the smallest non-zero value of 
$\widetilde\Delta$ such that $n_{BB}^{\widetilde\Delta}\geq1$, or 2. 
This result should hold for {\it any} CFT.
It is in fact related to the universal behaviour $\sim\exp\big({\rm const}\sqrt{c(EL-\pi c/6)}\big)$ \cite{JC86} of the density of states, 
valid for $EL\gg c$, after performing a steepest descent approximation for $\cal F$.
However, for  holographic CFTs, i.e. those with $c\gg1$ and a sparse density of states with $EL\ll c$, 
the above density of states formula is supposed \cite{holodensity} to hold also for $EL=O(c)$, in which case it may 
be argued that (\ref{eq3}) holds out to times $t=O(cL)$. 
Thus for these CFTs we expect to see no revivals at times $t=O(L)$. 
This is consistent with the holographic interpretation of the quench as forming a black hole in the AdS spacetime \cite{BH}.

On the other hand, for minimal CFTs $t=nL/2$ corresponds to $\tau\approx-n$. 
We may then relate the value of $Z_{\cal A}$ at this point to that near $\tau=0$, 
and then as $\tau\to i\infty$ using the transformation properties of the characters. 
In the limit $L/\tau_0\to\infty$, this gives
\be
{\cal F}(nL/2)=\sum_\Delta|B_\Delta|^2\left({\bf T}^n{\bf S}\right)_{\Delta,0}=\sum_{\widetilde\Delta}n_{BB}^{\widetilde\Delta}
\left({\bf S}{\bf T}^n{\bf S}\right)_{\widetilde\Delta,0}\,,
\ee
where $\bf S$ and $\bf T$ are the corresponding matrices according to which the characters transform. 
It follows that as long these are finite dimensional (as for the the minimal models or more generally a rational CFT), 
the value of ${\cal F}(t)$ at $t=nL/2$ is therefore finite (although, as we shall see below, it may accidentally vanish). 
At times within $(2L\tau_0)^{1/2}$ of this there is a similar decay to that in (\ref{eq3}) with $t$ replaced by $|t-nL/2|$. 
If $M$ is the lowest common denominator of all the $\{\Delta\}$, then, since all the energy gaps of $H$ (of even parity) 
are quantised in units of $4\pi/LM$, there must always be complete revival (${\cal F}=1$) at multiples of $t=ML/2$. 
For the minimal models, the Kac formula implies that in $M\sim24/(1-c)$ and therefore in general the time for a complete 
revival diverges as $c\to1^-$. 
We also find (numerically) that in the same limit the return amplitude at any fixed revival time goes to zero exponentially fast. 
A similar result should hold for other sequences of rational CFTs with a maximal value of $c$.

Although finite values of ${\cal F}$  occur only at integer values of $2t/L$, in fact there is interesting universal structure 
near every rational value. 
This is because the characters are singular at $\tau=0$, and the modular group maps this to every rational point 
$\tau=n/m$ on the real line.
This is mapped to $\tau=0$ by applying $ST^{n_1}ST^{n_2}\ldots$, where $(n_1,n_2,\ldots)$ are the integers 
appearing in the continued fraction expansion of $n/m$. However the nearby point $\tau=n/m+i2\tau_0/L$ 
is mapped to $\tau\approx im^2(2\tau_0/L)$ and so we find, after normalising with the denominator of (\ref{eq1}),
\be
{\cal F}(nL/2m)\propto (e^{-\pi L/\tau_0})^{(c/24)(1-1/m^2)}\,.
\label{eq4}
\ee
Once again, at nearby values of $t$, this is modified in a similar manner to (\ref{eq3}). 
A more careful analysis also shows that the correction terms may be neglected only for $m\ll(L/\tau_0)^{1/2}$, 
so that for a fixed $\tau_0/L$ the structure near only a finite number of rational values will be evident. 
This result shows that if we define a `large deviation function'
$-\lim_{L/\tau_0\to\infty}(\tau_0/\pi L)\log{\cal F}(t)$, it is a sum of delta functions of strength 
$\propto1/m^2$ at each rational value $n/m$ of $2t/L$, on top of the uniform plateau value $c/24$.
This structure may be understood in the quasiparticle picture as being due to the simultaneous 
emission at $t=0$ of entangled pairs of particles separated by distances which are integer divisors of $L$.

Many of these features are present in the simplest minimal CFT, corresponding to the scaling limit of the Ising model with $c=1/2$.
There are three distinct conformal boundary states, corresponding to the scaling limits of free and fixed boundary conditions on the 
Ising spins. In the last two cases \cite{JC89}, corresponding to a quench in the transverse field Ising model to the critical point 
from the ground state from the ordered phase,
\be
Z_{\cal A}^{\rm fixed}=\frac12\chi_0(q)+\frac12\chi_{1/2}(q)+\frac1{\sqrt2}\chi_{1/16}(q)=\chi_0(\tilde q)\,.
\ee
At the recurrence times $t=nL/2$, we find by applying ${\bf T}^{n}$ and then $\bf S$
\be
\fl Z_{\cal A}^{\rm fixed}=
\frac12\chi_0(q')+\frac12e^{i\pi n}\chi_{1/2}(q')+\frac1{\sqrt2}e^{i\pi n/8}\chi_{1/16}(q)
\simeq \left[\frac14(1+e^{i\pi n})+\frac12e^{i\pi n/8}\right]\chi_0(\tilde{q'}),
\ee
where $q'=e^{-4\pi\tau_0/L}$, $\tilde{q'}=e^{-\pi L/\tau_0}$, and we have retained only the dominant term in the second step.
For $n$ odd this gives ${\cal F}(nL/2)=1/2$, while for $n$ even we get $|\cos(\pi n/16)|$. 
There is complete revival at $t=8L$, while at $t=4L$ the coefficient vanishes, leaving a much smaller term 
$O((e^{-\pi L/\tau_0})^{1/16})$. 

\begin{figure}
\includegraphics[width=0.45\textwidth]{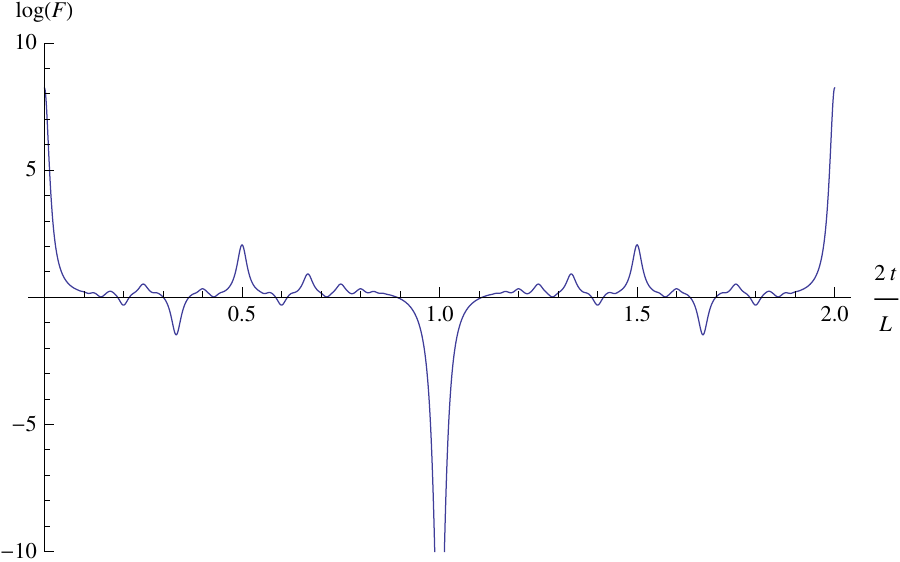}
\includegraphics[width=0.45\textwidth]{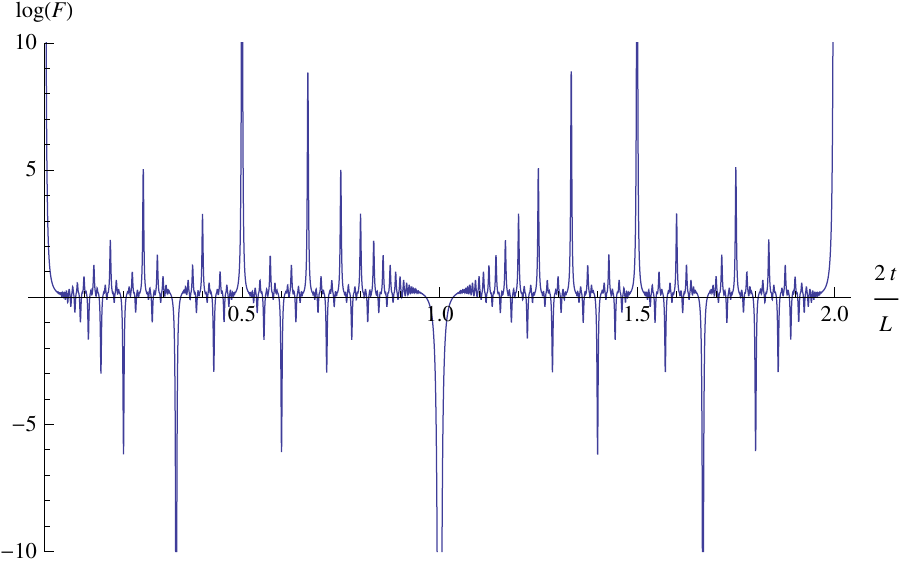}
\caption{\label{fig2}
Left: Log of the return amplitude for the Ising CFT starting from a disordered state for $0<2t/L<2$, with $\pi\tau_0/L=0.05$. 
The vertical axis has been shifted so as to expose the mean plateau behaviour. 
This shows the initial gaussian decay and revival at $t=L$. The negative peak at $t=L/2$ is due to destructive interference between two kinds of quasiparticles. Smaller gaussian peaks are seen at rational values with small denominators.
The positive peaks are mapped by the modular group to the initial peak, and the negative ones to the feature at $2t/L=1$.
Right: Same as left with $\pi\tau_0/L=0.005$. 
Now there is structure at more rational values, and we see the predicted $1/m^2$ 
dependence of the heights of nearby peaks with denominators $m$.
Reprinted with permission from Ref. \cite{c-14}
}
\end{figure}

On the other hand, for free boundary conditions \cite{JC89}, corresponding to a quench from the disordered phase,
\be
Z_{\cal A}^{\rm free}=\chi_0(q)+\chi_{1/2}(q)=\chi_0(\tilde q)+\chi_{1/2}(\tilde q)\,.
\ee
At $t=nL/2$, we get $\chi_0(q')+(-1)^n\chi_{1/2}(q')$, so for $n$ even there is complete revival, but, for odd $n$,
$\chi_0-\chi_{1/2}=\sqrt2\chi_{1/16}(\tilde{q'})$, so again the revival is suppressed. The above expression may also be written as 
$Z_{\cal A}^{\rm free}=q^{-1/48}\prod_{k=0}^\infty(1+q^{k+1/2})$, which explicitly shows the structure near rational values of $2t/L$. 
This is illustrated in Fig.~\ref{fig2}.

\subsection{Quench from a more general boundary state and the GGE}
\label{genIS}

So far we have considered a quench from a state of the form $e^{-\tau_0H}|\psi_0^*\rangle$. 
This was chosen mainly for the fact that it leads to simple analytic results, but it also has the property of 
thermalisation to a Gibbs distribution for any subsystem. 
Instead, in an integrable model, as a consequence of the existence of an infinite number of conserved quantities commuting 
with the hamiltonian, the stationary state is expected to be described by a generalised Gibbs ensemble (GGE) rather than a  
Gibbs distribution \cite{GGE}. 
This has been shown to be correct for models admitting a free-particle representation, as \em e.g. \em in 
Refs.~\cite{GGE,cc-07,cef-11,cef-12b,fe-13,bkc-14,Caza,bs-08,cdeo-08,scc-09,cic-12,csc-13,sc-14,ga-15,pe-16}.
Some GGEs were then explicitly constructed for truly interacting integrable models for which 
an exact solution for the stationary state was not yet available \cite{fm-10,mc-12b,ck-12,ks-13,fe-13b,bp-13a,fcec-14,a-15}.
It has later been shown that the GGE built with the (ultra-)local charges were correct only for those models with 
a single species of quasi-particles \cite{kcc-14,stm-14,bp-14,dwbc-14,bpc-16}.
Contrarily, in order to match the exact solutions for more general models \cite{amsterdam,budapest,pce-16},
ultra local charges are not sufficient \cite{f-14b,ga-14,bp-14b,emp-15,f-16}.
Only recently it has been understood  \cite{idw-15} that newly discovered quasi-local charges 
\cite{ppsa-14,mpp-15,pv-16,impz-16} must be indeed included in the GGE to match the exact solutions. 
All these results in integrable models rise a natural question: 
How a GGE can appear in a CFT, which (in the rational case) is the most integrable model of all? 
In this section we argue that the relaxation to the Gibbs distribution in a CFT is a consequence of the choice of initial state
(\ref{tau0B}), and that a more general state leads to a form of the GGE. 
The arguments are a condensed version of those given in \cite{c-15}.

We remind the reader that boundary renormalisation group (RG) shows that each bulk CFT has a particular allowed set of 
boundary states, each of which corresponds to a fixed point of the RG with its own basins of attraction. 
For the minimal models, there is a finite number of these boundary states, and their basins of attraction are expected 
to contain the ground states of all non-critical hamiltonians $H_0$. 
Thus this ground state should be representable in terms of all possible irrelevant operators acting on $|\psi_0^*\rangle$
\be
|\psi_0\rangle\propto \prod_ke^{-\beta_k\int\widetilde\Phi_k dx}|\psi_0^*\rangle\,,
\label{generalstate}
\ee
suitably regularised. 
One of these irrelevant operators is the component $T_{tt}$ of the stress tensor, whose space integral is the hamiltonian. 
Thus the ansatz (\ref{tau0B})  is equivalent to assume that this is the  only term in the above sum. 
$T_{tt}$ may be written as a sum of holomorphic and antiholomorphic operators $T$ and $\overline T$, 
each of which is conserved and whose space integral is a conserved charge. 
This is true also for all the irrelevant operators which are its descendants, that can be written as powers 
of $T$ and derivatives thereof, plus their antiholomorphic partners. 
As we will show,  the conformal quench dynamics starting from such a state 
leads to a stationary reduced density matrix for a finite interval $A$ which has the GGE form
\be
\rho_A\propto {\rm Tr}_{\overline A}\,\prod_ke^{-\beta_k\cos[\pi\Delta_k/2]\int(\Phi_k+\overline\Phi_k) dx}\,,
\label{GGE}
\ee
where  now $\Phi_k$ and $\overline\Phi_k$ are  holomorphic and antiholomorphic bulk operators so 
that their space integrals are conserved charges. 
Notice, that the charges 
\be
H^{(k)}=\int \big(\Phi_k(x,\tau)+ \overline\Phi_k(x,\tau)\big)dx\,,
\label{Hk}
\ee
do not necessarily commute among themselves, but they all commute with the CFT hamiltonian. 
The standard GGE includes only a commuting sub-algebra of the $H^{(k)}$. 
This is motivated by the idea that these should form a complete set of  commuting observables which should 
characterise any macrostate. 
However, this does not seem to be the case for a 1D CFT for which, because of the exactly linear dispersion relation, 
there is a massive degeneracy of states, and the expectation values of the charges in the commuting sub-algebra 
(identified in \cite{Zam}) are not sufficient to characterise the states.  
A similar phenomenon has been pointed out by Sotiriadis \cite{Sot} in the context of a quench in a 1D massless 
free boson theory from a non-gaussian initial state, although in this case the commuting set of conserved charges 
is formed by the mode occupation numbers.

However, this treatment still does not account for all possible irrelevant boundary operators in the state (\ref{generalstate}). 
In \cite{c-15} it is  argued that, at least in rational CFTs, for every boundary operator $\widetilde\Phi_k$ there is a pair of holomorphic 
and antiholomorphic operators $(\Phi_k,\overline\Phi_k)$ with the same overall scaling dimension $\Delta_k$. 
These also lead to conserved charges and a generalisation of (\ref{GGE}).
However, since $\Delta_k$ can be non-integer, these currents are parafermionic and the corresponding charges are 
non-local. 
Although such charges are not customarily included in the GGE, they make perfect sense
and they lead to physically sensible results.

Let us rewrite the initial state (\ref{generalstate}) as
\be
|\psi_0\rangle\propto e^{-2\tau_0H}\,\prod_ke^{-\beta_k\int\widetilde\Phi_k(x)dx}|\psi_0^*\rangle\,,
\label{eqphib}
\ee
where the product is now over all boundary operators except the stress tensor. 
We have commuted $e^{-2\tau_0H}$ through all the other terms so that it stands on the left, but  
since commutators of $H$ with a given local operator generate others, the net effect is just to modify the values of $\beta_k$.  

As it stands, Eq.~(\ref{eqphib}) is only formal because, expanded in powers of the $\beta_k$, it involves integrals of boundary
correlators of the $\widetilde\Phi_k$ which are UV divergent if $\Delta_k>1$. 
These divergences may be canceled in the usual way by imposing a UV cut-off in real space, 
identifying the UV divergent terms by using the OPE, and  adding counterterms. 
This has the effect that the renormalised $\beta_k$ appearing in (\ref{eqphib}) then depend on the bare 
values in a complicated non-linear fashion. 

This perturbative expansion is expected to make sense for all irrelevant boundary operators with $\Delta_k>1$. 
For relevant operators one may expect to encounter infrared divergences signalling the crossover to a different boundary fixed point. 
However, $\tau_0$ provides a cut-off, and the expansion should 
still make sense as long as $\beta_k^{-1/(1-\Delta_k)}\gg\tau_0$, i.e. $\beta_k\ll\tau_0^{-(1-\Delta_k)}$.

The argument is (at least formally) quite simple. On the boundary, $\widetilde\Phi_k=\Phi_k=\overline\Phi_k$, so we may write
\be
\beta_k\int \widetilde\Phi_k(x)dx=\frac{\beta_k}2\int \Phi_k(w)dw+\frac{\beta_k}2\int\overline\Phi_k(\bar w)d\bar w,
\ee
along contours just above and below the lower and upper boundaries $\tau=\pm\tau_0/2$. 
There is a contour for each $k$, and they should be ordered in increasing $k$. 
Because of the (anti-)holomorphicity, each of these contours may be freely deformed into the bulk, 
as long as they do not cross each other or the arguments of local observables in correlation functions. 
This corresponds to the statement that the charges which are the space integral these currents commute 
with the hamiltonian (up to boundary conditions in the case of fractional spin -- see later.) 
On transforming to the $z$-plane, the insertion becomes (for each $k$)
\be
\frac{\beta_k}2(-2i\tau_0/\pi))^{1-\Delta_k}\int \Phi_k(z)z^{\Delta_k-1}dz+ \cdots +{\rm c.c.}\,,
\ee
summed over two contours: one from $z=0$ to real $+\infty$, the other from $z=0$ to real $-\infty$. 
The omitted terms in the above are the contributions of other (holomorphic) descendants, 
which are present because some of the $\Phi_k$ are not primary. 
These will disappear when reversing the mapping going to the cylinder.

Let us now continue the arguments of the local observables to real time. 
At late times the $\bar z_j$ move off to $-i\infty$ so that effectively the boundary disappears. 
However, the boundary perturbations given by the contour integrals remain. 
In the absence of the boundary, we can rotate one contour into the other, giving a relative factor
$(e^{-i\pi/2})^{\Delta_k}+(e^{i\pi/2})^{\Delta_k}=2\cos(\pi\Delta_k/2)$.
Only even integer dimensional charges contribute because the odd ones are odd under parity and vanish on the initial state.

On transforming back to the $w$-plane (which is a cylinder), we find an insertion in the path integral action
of the term $\sum_k\beta_k\cos(\pi\Delta_k/2)H^{(k)}$ along $\tau=\tau_0$. 
However, given that all $H^{(k)}$ commute with the hamiltonian, this contour could be along any constant imaginary time 
line that does not separate any of the arguments $w_j$ of the operators. 
However, the contours for different $k$ must be correctly ordered, as a consequence of  the fact that the $H^{(k)}$ 
do not in general commute among themselves. 

Thus the reduced density matrix (at times after all the subsystem has fallen within the same horizon) 
are given by a path integral with weight
\be
e^{-S_{CFT}-\sum_k\beta_k\cos(\pi\Delta_k/2)H^{(k)}}\,,
\label{SGGE}
\ee
where the first term is the standard CFT action. Eq.~(\ref{SGGE}) is the desired path integral formulation of a (non-Abelian) GGE.

This CFT GGE has many observable consequences, described in detail in \cite{c-15}. 
The main feature is that the effective inverse temperature, of which there is just a single one  $\beta$ in the Gibbs ensemble, now depends on the observable -- for example, it is different for each equal-time correlation function $\langle\Phi\Phi\rangle$ if extracted from the correlation length $2\pi x_\Phi/\beta_{\rm eff}$, an effect which also shows up in the time-dependence of the 1-point function $\langle\Phi(r,t)\rangle$; it gives a different relation between the entropy and the energy density, and so on. 

These somewhat formal results may be checked by perturbation theory in the $\beta_k$. For example, the first order correction to the exponential decay of a 1-point function gets a relative correction $O(\beta_kt)$. This may be seen to come from the boundary operators within a distance $t$ of the point $r$. Although this appears to be larger than the zeroth order term, it may be argued that it then exponentiates to give the required $O(\beta_k)$ correction to the decay rate. This has been verified for the Ising model \cite{cef-11} with an explicit form for the decay rate.

When the $\Delta_k$ are non-integer, however, 
the associated charges $H^{(k)}$ (\em cf. \em Eq.~(\ref{Hk})) do not quite commute with the hamiltonian.
 This is because if we consider a closed contour which is the boundary of a long rectangle $(-L/2<u<L/2,\tau_1<\tau<\tau_2)$, the contributions from the end pieces do not cancel for periodic boundary conditions since they carry different phases $e^{\pm i\Delta_k}$. They would in fact cancel for suitable twisted boundary conditions on the fields of the theory. Thus we may think of the action of a single $H^{(k)}$ as switching the boundary conditions between different sectors of the theory. It is a parafermionic charge. 
 If $\Delta_k$ is rational, with lowest denominator $M$, it is only after acting $M$ times that we return to the original boundary conditions. Thus, in the GGE expression
\be
{\rm Tr}\, e^{-\beta H }\prod_ke^{-\beta_k\cos(\pi\Delta_k/2) H^{(k)}}\,,
\ee
the trace  projects onto only those terms in the expansion of the exponential containing integer powers of $\beta_k^M$. 

As an example, one can consider a quench in the Ising chain with both transverse and longitudinal field, i.e. 
hamiltonian (\ref{Hamil}) with the addition of $-h_x\sum_j\sigma_j^x$.
One starts from the ground state with $h_z\gg 1$ and $|h_x|\ll 1$ and quench to the critical point $h_z=1$, $h_x=0$.
The appropriate conformal boundary state when $h_x\gg 1$ and $h_z=0$ is that corresponding to free boundary conditions 
on the $\sigma_j^x$. This state supports one primary operator of scaling dimension $\delta=1/2$ 
which is interpreted as the scaling limit of the local magnetisation $\sigma^x$. 
Turning on a small $h_x$ is equivalent to perturb the boundary state with a relevant operator and so we
 we have to consider the window $h_x\ll \beta^{-1/2}\sim(h_x-1)^{1/2}$, as argued above. 
The holomorphic and anti-holomorphic extensions of this boundary operator are in this case well understood 
-- they are nothing but the fermions $(\psi(z),\bar\psi(\bar z))$ of the bulk Ising CFT. 
Thus the GGE in the case should contain fermionic charges which, as is well known, 
act to switch between periodic and anti-periodic boundary conditions on the Ising spins $\sigma^z$. 

Finally, it must be mentioned that these results are also relevant to the physics of prethermalization
\cite{kehrein,exp-pt,kwe-11,mmgs-13,ekmr-14,bck-14,nic-14,bf-15,fc-15,begr-15} in models with weak integrability breaking. 
A prethermalized regime has been observed in experiments \cite{exp-pt} and it is a crossover from a 
prethermalization plateau described by a GGE (as in this section) to the truly stationary thermal state of the 
non-integrable model.

\subsection{Quench to a CFT perturbed by an irrelevant operator}

We consider what happens when the time evolution is governed by a hamiltonian differing from that of a CFT by 
the addition of irrelevant operators. In equilibrium, these give only corrections to scaling, but their influence 
on the quench dynamics is less obvious. 

It is clearly possible to study these effects in perturbation theory of the couplings to the irrelevant operators, as in 
the case of the deformed initial state above. This perturbation theory breaks down at late times. 
Here we take a different approach, which agrees with perturbation theory at low orders. 
However, this approach  works only for operators that are descendants of the identity, \em i.e.~\em polynomials of the stress tensor 
components and their derivatives. 
Here, for simplicity, we restrict to considering the dimension 4 operators $T^2$, $ \overline T^2$ and $T\overline T$. 
These are generated in almost any quantum critical hamiltonian.
$T^2$, $ \overline T^2$ break Lorentz invariance, they are always generated in a lattice model,
their space integrals commute with the hamiltonian, and they preserve the integrability of the model. 
 $T\overline T$ preserves relativistic invariance, but introduces right-left scattering and does 
not commute with the conformal hamiltonian, and generally makes the theory non-integrable.

We then consider the  hamiltonian
\be
H=H_{CFT}-\frac12\lambda\int(T^2+\overline T^2)dx-\mu\int T\overline Tdx\,,
\ee
corresponding to the euclidean action
\be
S^E=S^E_{CFT}-\frac12\lambda\int(T^2+\overline T^2)dxd\tau-\mu\int T\overline Tdxd\tau\,.
\ee
The additional terms can be generated by the standard trick of introducing auxiliary  fields 
$(\bar\xi,\xi)$, with spins $\pm2$, in the functional integral
\be
\fl S^E=S^E_{CFT}+\frac1{\mu^2-\lambda^2}\int\big(\mu\bar\xi\xi-\frac12\lambda(\bar\xi^2+\xi^2)\big)dxd\tau+\int(\bar\xi T+\xi\overline T)dxd\tau.
\label{SE}
\ee
In Feynman diagram language, this interaction is an exchange of $(\bar\xi,\xi)$ `particles' which couple linearly 
to $T$ and $\overline T$. 
A UV regulator can be introduced in the form of a kinetic term $\propto\int(\partial\xi)(\bar\partial\bar\xi)dxd\tau$. 
The couplings $\lambda,\mu$ have dimension (length)$^2$. We assume that they are small compared to the dominant 
length scale of the problem $\tau_0^2$ so that we can assume that the (dimensionless) fields $(\bar\xi,\xi)$ are $\ll1$. 

The last term in the action (\ref{SE}) may be reinterpreted as the response to a small change in the metric 
$\delta\bar g=\delta g^{ww}=\bar\xi$, $\delta g=\delta g^{\bar w\bar w}=\xi$,
which themselves may be viewed as gaussian random fields with covariance
\bea
{\mathbb E}[\delta g(u,\tau)\delta g(u',\tau')]&=&{\mathbb E}[\delta\bar g(u,\tau)\delta\bar g(u',\tau')]=\lambda\delta(u-u',\tau-\tau')\,,
\nonumber \\
{\mathbb E}[\delta g(u,\tau)\delta\bar g(u',\tau')]&=&\mu\delta(u-u',\tau-\tau')\,.
\eea
The deformed metric is
\be
ds^2=dwd\bar w+\delta gdw^2+\delta\bar gd\bar w^2\,,
\ee
which to lowest order is equivalent to the coordinate change
$ds^2=d\tilde wd\bar{\tilde w}$ with
\be
d\tilde w=dw+\delta \bar gd\bar w,\quad d\bar{\tilde w}=d\bar w+\delta gdw\,.
\ee
This implies, to lowest order, that the correlation functions of any product of local observables evaluated with the action (\ref{SE}) 
is the expectation value over the random fields of the same product in the pure CFT at shifted values of the arguments
\be
\langle{\cal O}(w_j,\bar w_j)\rangle_{S_{CFT}+\delta S}=
{\mathbb E}[\langle{\cal O}(\tilde w_j,\bar{\tilde w}_j)\rangle_{S_{CFT}}]\,.
\ee
where ${\mathbb E}[\cdot]$ stands for the average over the gaussian fields.

We are finally ready to apply these ideas to the quench problem. 
Given that a point $(x_j,t_j)$ in Minkowski space is mapped to
$w_j=x_j-t_j$, $\bar w_j=w_j'=x_j+t_j+2i\tau_0$, on the cylinder, the effect of the coordinate shift in light-cone coordinates is
\be
d\tilde x_+=dx_++\delta\bar g\,dx_-\,,\quad d\tilde x_-=dx_-+\delta g\,dx_+\,.
\ee

In the shifted coordinates the correlations have to be evaluated in the pure CFT, in which signals propagate with speed $v=1$. 
The left- and right-moving null geodesics are  $d\tilde x_+=0$ and
$d\tilde x_-=0$, that the original coordinates read
\be
dx_+=-\delta\bar g\,dx_-\,,\qquad dx_-=-\delta g dx_+\,.
\label{dx}
\ee

There are several consequences of this change of coordinate. The first one  is that,  
given that $\langle T\rangle$ and $\langle\overline T\rangle$ are both non-vanishing, 
$\delta g$ and $\delta\bar g$ have non-zero expectation values   
(from (\ref{SE}) at the saddle-point $\langle\delta g\rangle=\langle\xi\rangle=-(\mu+\lambda)\langle T\rangle\not=0$).
Neglecting fluctuations, the null geodesics are $dt=\pm(1+2\langle\delta g\rangle)dx$, so the average velocity 
of propagation is renormalised.
Since $\langle T\rangle=\langle\overline T\rangle<0$ in the initial state (and thereafter), $\langle\delta g\rangle$ 
and $\langle\delta\bar g\rangle$ are both positive if $\mu+\lambda>0$, and so the speed is reduced. 
It may be argued that this corresponds to an attractive interaction.

The second effect is a diffusive broadening of the horizon. In order to show this, we need to 
include the  random nature of $(\delta g,\delta\bar g)$ in Eq. ~(\ref{dx}). 
Integrating (\ref{dx}) to lowest order in the couplings, the equations for a right-moving null geodesic starting at $(x_+=0,x_-=x_-(0))$ is
\be
x_-(x_+)=x_-(0)-\int_0^{x_+}\delta g(x_+',x_-(0))dx_+'\,.
\ee
In order to properly cure the UV behaviour, 
we have to use the fact that the horizon in the pure CFT has a width $O(\tau_0)$. 
Defining then the mean $x_-$ coordinate of the horizon by the random variable
\be
X_-(x_+)=(4\tau_0)^{-1}\int_{-2\tau_0}^{2\tau_0}x_-(x_+)dx_-(0)\,,
\ee
this has variance
\be
\langle X_-(x_+)^2\rangle=(\lambda/4\tau_0)x_+\,.
\ee
This shows that the effect of $T^2$ and $\overline T^2$  is to diffusively broaden the horizon
to a width $O\big((\lambda t/\tau_0)^{1/2}\big)$. 
However, since  this is smaller than $t$, the horizon remains well-defined. 
The $T\overline T$ term does not contribute to this effect.

It can be shown that this effectively fluctuating metric leads to modification of correlation lengths, consistent with that found from perturbation theory to lowest order. In a finite volume, it also leads to a progressive broadening and attenuation of the 
revivals (\em cf. \em Sec.~\ref{sec:rev}).

We finally discuss the effects of higher orders in $\lambda$ and $\mu$, as well as higher order descendants $T^{(k)}$ 
of the stress tensor (this can be described by a coupling to a random metric with non-gaussian correlations, that are not 
expected to change the overall picture when they are small). 
It is always possible to choose a local system of coordinates $\tilde x_\pm$ so that the metric is
\be
ds^2=e^{\phi(\tilde x_+,\tilde x_-)}\,d\tilde x_+d\tilde x_-\,.
\ee
Starting from the second order, this corresponds to non-zero curvature  $R\propto\partial^+\partial^-(\delta g_+\delta g_-)$, 
that appears as a consequence of left-right scattering. 
In fact, LL and RR scatterings are equivalent to a coordinate change $x_\pm\to\tilde x_\pm$ and cannot introduce a curvature  
leading only to broadening of the horizon, possibly non-gaussian. 
Instead, strong negative curvature effects might lead to other effects such as the chaotic divergence of geodesics 
which are very difficult to analyse quantitatively. 

It should be interesting to extend our analysis of the non-integrable $T\overline T$ perturbation of the CFT to other situations, for example an inhomogeneous quench (\em cf. \em Sec.~\ref{sec:inoh}), where there are non-zero energy currents, or to the steady state currents set up by a non-zero temperature gradient (see \cite{BD} and \cite{transp}). 
Recently this problem has been addressed from a different point of view \cite{BD2}.

\section{CFT approach to local quantum quenches} 
\label{sec:local}

In a local quantum quench the initial state differs only locally (i.e. 
in a small finite region) from the ground state of the model. 
Although there is such a small (indeed infinitesimal compared to the ground state energy) excess of energy 
in the  initial state, this has a very large effect which may be 
interpreted as a manifestation of the famous Anderson orthogonality catastrophe \cite{a-67}.
Obviously the details of the many-body dynamics depend on the way the local perturbation to the 
ground state is performed. 
One case is treated by means of CFT in a particularly simple fashion \cite{cc-07l} and it is the so called 
{\it cut and glue} protocol. 
It works as follows: we physically cut a one dimensional quantum system (\em e.g. \em a spin chain) 
at the boundaries between two subsystems $A$ and $B$, and prepare a state where the two individual
pieces are in their respective ground states. 
In this state $A$ and $B$ are unentangled, and its energy differs from that of the ground state by only a finite amount. 
At time $-t$ we glue the two pieces  and let the system evolve up to $t=0$. 
Clearly the method outlined previously for global quenches does not apply because the
initial state is not translational invariant.

\begin{SCfigure}%[t]
\includegraphics[width=7cm]{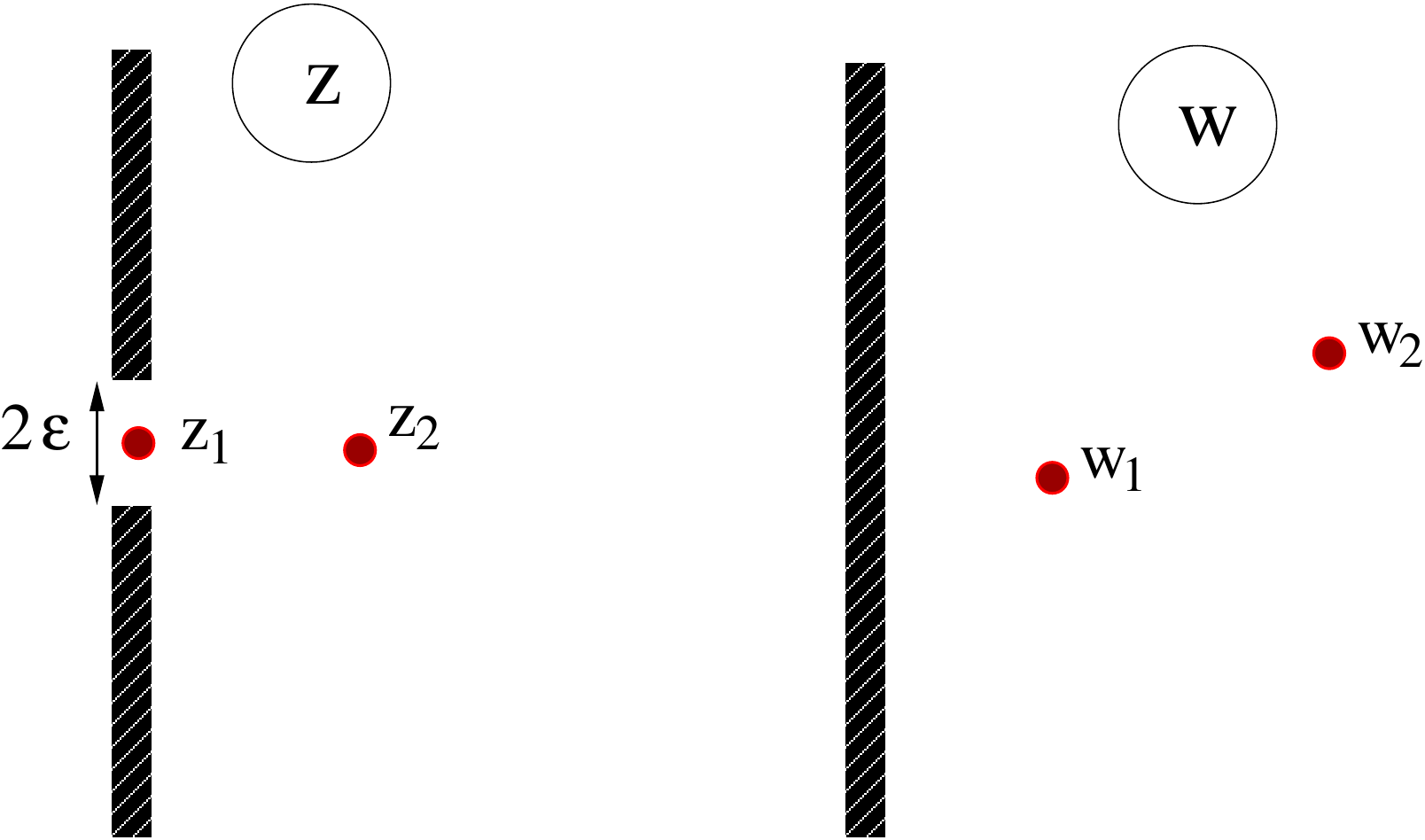}
\caption{Space-time region for the density matrix for the local
quench (left) mapped to the half-plane (right) by means of
Eq.~(\ref{mapp1}). $z_1=i\tau$ and $z_2=i\tau+\ell$. 
Reprinted with permission from \cite{cc-07l}.} \label{map1}
\end{SCfigure}

There is however an alternative and convenient way to represent the corresponding density matrix in terms of a path integral. 
The physical cut of the pre-quench hamiltonian corresponds to having two  slits parallel to the imaginary time axis, 
one starting from $-\infty$ up to $\tau_1=-\epsilon-i t$ (the time when the two pieces have been joined), 
and another from $\tau_2=\epsilon-i t$ to $+\infty$.
This plane with the two slits is shown on the left of Fig.~\ref{map1}.
We introduced the regularisation factor $\epsilon$ in order to have a finite distance between the two 
slits (in analogy to the requirement of a finite strip in the global quench).
The physical operators are inserted at imaginary time $\tau=0$.  
However, for computational simplicity we consider the translated geometry (by $it$) with two cuts starting at $\pm i\epsilon$ and
operators inserted at imaginary time $\tau$. 
This should be considered real during the course of the computation, and only
at the end can be analytically continued to $it$. 
As shown on the right of Fig.~\ref{map1}, the $z$-plane is mapped into the 
half-plane ${\rm Re}\, w>0$ by the conformal mapping 
\be
w=\frac{z}{\epsilon}+\sqrt{\left(\frac{z}{\epsilon}\right)^2+1} 
 \label{mapp1}
\ee 
On the two slits in the $z$ plane (and so on the imaginary
axis in the $w$ one) appropriate conformal boundary conditions compatible with the cut must be imposed 
(\em e.g. \em Dirichlet for a free boson).

As already stated, in a local quantum quench, there is only a small excess of energy compared to the ground-state. 
Consequently, only low lying excited states are expected to be populated after the quench.
When a quantum system such as a spin chain is  effectively described at low energy and in equilibrium by a CFT, we then 
expect that even the dynamics following a quantum quench is correctly captured by CFT when time and distances 
are much larger than any microscopical scale. This includes also the regulator $\epsilon$ introduced above.
Thus, as a fundamental difference compared to global quenches, we expect the results for local quenches to be {\it universal} 
and valid for any system whose low energy spectrum is captured by CFT.

\subsection{Entanglement between the two halves} 

The first question we consider is how the entanglement entropy between the two parts in which the 
system was divided before the quench grows in time.
As we shall see, this is obtained by a very simple calculation and gives very important information about the quench. 
As already discussed in Sec.~\ref{entaglob}, the moments of the reduced density matrix $\Tr \rho_A^n$ transform 
like a one-point function of twist operator of scaling dimension $x_n$ given by Eq.~(\ref{twistx}).
In the $w$ plane this one-point function is $\tilde{c}_n[2{\rm Re} w_1/a]^{-x_n}$ (with $\tilde{c}_n$  non-universal constants 
and $a$ the UV cutoff such as a lattice spacing). 
Thus in the $z$ plane at the point $z_1=(0,i\tau)$ we have
\be
\langle\tw_n(z_1) \rangle=
\tilde{c}_n \left( \left|\frac{dw}{dz}\right|_{z_1}
\frac{a}{[2{\rm Re} w_1]}\right)^{x_n}=
\tilde{c}_n\left(\frac{a\epsilon/2}{\epsilon^2-\tau^2}\right)^{x_n}\,,
\label{phin}
\ee
that continued to real time $\tau\to it$ is
\be
\langle\tw_n(t)\rangle=\tilde{c}_n\left(\frac{a\epsilon/2}{\epsilon^2+t^2}\right)^{x_n}\,.
\ee
Using finally the replica trick to find the entanglement
entropy we obtain
\be
S_A=- \left.\frac{\partial }{\partial n}\Tr
\rho_A^n\right|_{n=1}= \frac{c}{6}\ln \frac{t^2+\epsilon^2}{a\epsilon/2}+\tilde{c}'_1
\simeq
\frac{c}{3}\ln \frac{t}a +{\rm const}\,, \label{Slogt}
\ee
i.e. the leading long time behaviour is only determined by the central charge of the theory.
Eq.~(\ref{Slogt}) is reminiscent of the ground state result for an interval of length $\ell$ \cite{cc-04,cc-09}, 
but the prefactor $c/3$ has a different origin since in this case there is only one boundary point between the two 
subsystems $A$ and $B$. Indeed there is a factor two compared to what one could naively expect and this follows 
from having introduced an new length scale $\epsilon$.

This long-time $\ln t$ behaviour has been analytically and numerically observed many different models in the 
cut and glue scenario \cite{gkss-05,ep-07,ekpp-07,kkmgs-07,hgf-09,isl-09}.
Furthermore the $\ln t$ growth is more generally valid also when the two halves are not joined in a translational 
invariant manner.  
For example for free fermionic model joined with a bond defect there is a logarithmic growth with a 
renormalised prefactor \cite{ep-12,eg-10}, while in interacting models the prefactor depends upon the relevance (in RG sense) 
of the impurity bond \cite{cc-13}.

\subsection{Entanglement of a de-centered subsystem}

We now consider the entanglement entropy of the region $r>\ell$ with
the rest of the system. In this case $\Tr \rho_A^n$ is equivalent
to the one-point function in the plane $z$ at the point $z_2=\ell+i\tau$ (see  Fig.~\ref{map1}). 
Using the conformal
mapping (\ref{map1}), analytically continuing and taking
$t,\ell\gg \epsilon$ one finally get \cite{cc-07} 
\be S_A= \cases{
\frac{c}6 \ln \frac{2\ell}a + \tilde{c}'_1  & $t<\ell$\,,\cr
\frac{c}6 \ln \frac{t^2-\ell^2}{a^2} + {\rm const} & $t>\ell$\,. } 
\label{s22}
\ee
The interpretation of this result is direct. Quasi-particle excitations produced by the gluing at $r=0$ take a time $t=\ell$ to
arrive at the boundary between $A$ and $B$ and only at that time start modifying their entanglement. 
Note that for $t\gg\ell$ we recover Eq.~(\ref{Slogt}).
The time independent value for $t<\ell$  is the entanglement of a finite interval of length $\ell$ 
 in the half-line \cite{cc-04}. The singularity at $t=\ell$ is smoothed on a scale $\epsilon$.

In Ref. \cite{cc-07l} also the entanglement of a finite interval (both on one side or across the cutting point) 
has been derived. We do not review these results here, but refer the interested reader to the original 
literature. Further results for more complicated bipartitions have been derived in \cite{ab-15}.
Also the entanglement between non complementary parts (quantified by the negativity) 
has been considered in \cite{ez-14,wcr-15}.

\subsection{One-point function of a primary operator}

The one-point function of a primary field in the half-plane
 ${\rm Re}\, w>0$ is $\langle \Phi(w) \rangle= {A^\Phi_b}{[2 {\rm Re}\, w]^{-x_\Phi}}$,
where $x_\Phi$ is the scaling dimension of the field and $A^\Phi_b$
is the same amplitude in appearing Sec.~\ref{1pt-glob}.
In this section, we fix the UV cutoff $a$ to $1$. 
Using the mapping (\ref{mapp1}) we can obtain the time dependence of  the one-point function of a primary operator. 
At the point $r$, after continuing to real time we have 
\be \langle \Phi(r,t) \rangle= \cases{ A^\Phi_b (2
r)^{-x_\Phi} &\qquad $t<r$\,, \cr A^\Phi_b \left(\frac{\epsilon}{2(t^2-
r^2)}\right)^{x_\Phi} &\qquad $t>r$\,. } 
\ee 
Thus for short times the correlation takes its initial value, until the effect of the
joining arrives at time $t=r$ when it decays for $t\gg r$ like
$t^{-2x_\Phi}$ (note that this exponent is twice the boundary one).
This prediction has been explicitly verified in free fermionic models \cite{dir-11,irt-14}.

\subsection{Two point function of a primary operator}

In order to calculate the two-point function of primaries $\langle \Phi(r_1,t) \Phi(r_2,t)\rangle$
after a global quantum quench, one should start from the 
general scaling form for this correlator in the  half-plane (\ref{2ptgen}), use the map (\ref{mapp1}) and finally take 
the limit $t,r_1,r_2,r_{12}\gg \epsilon$.
The algebra is quite long, but elementary. We then refer for details to the original literature \cite{cc-07l}
and we  report only the result here. 

There are three regimes for the time dependence of this correlation. 
Without loss of generality we can assume $r_1>0$ and $|r_2| <r_1$. 
For large times $t>r_1$ (and hence $t>|r_2|$) we have 
\be
\langle \Phi(r_1,t) \Phi(r_2,t)\rangle= \frac1{|r_1-r_2|^{2 x_\Phi}}\,,
\ee
i.e. the correlation function relaxes to its ground-state value.
This can be understood by the fact that the small excess of energy introduced by the defect in 
the quasi-particle picture is pushed far away for large time. 

The behaviour for $|r_2| <t<r_1 $ is more complicated and can be  written as 
\be\fl
\langle \Phi(r_1,t) \Phi(r_2,t)\rangle=\left[
\frac{(r_1 + r_2)(r_2+t)}{(r_1 - r_2) (r_1 -t)} \frac\epsilon{4r_1 (t^2 -
r_2^2)}\right]^{x_\Phi}
F\left( \frac{2 r_1 (r_2 + t)}{(r_1 + r_2)(r_1 + t)}\right) \,.
\ee
For $t<|r_2|$ we see that the gluing did not reach any of the two points 
and the correlation keeps its initial value which depends on whether the two points are on the same or on different side 
of the cutting point. Also this prediction for the two-point function has been explicitly verified in free fermionic models \cite{dir-11}.

\subsection{The return amplitude}

As firstly pointed out in \cite{ds-10}, the return amplitude\footnote{This is sometimes (incorrectly) referred to as the Loschmidt echo.} $\langle \psi_0 |\psi(t) \rangle$
after a local quantum quench in a CFT display a universal behaviour for large time.  
Let us consider 
\be
{\cal F}(t)\equiv- \ln |\langle \psi_0 |\psi(t) \rangle|=-\ln |\langle \psi_0|e^{-iHt} |\psi_0 \rangle|,
\label{LE}
\ee
sometimes also called bipartite logarithmic fidelity \cite{ds-10}.
In imaginary time (and having introduced the regulator $\epsilon$), ${\cal F}(\tau)$ is nothing but 
the free energy of the statistical mechanical system defined in the geometry shown in left of Fig.~\ref{map1}.
In a CFT this is obtained by computing the expectation value of the stress energy tensor in this geometry
(which is easily determined by using its transformation \cite{bpz-84} under the conformal mapping (\ref{mapp1})) that is the
derivative of the free energy wrt $\tau$.
The desired free energy is then obtained by integrating this expression. 
The details of this calculation are reported in \cite{ds-10,ds-11} with final result
\be
{\cal F}(\tau)= \frac{c}4\ln |\tau|+{\rm cnst}\,,
\ee 
that continued to real time yields 
\be
{\cal F}(t)= \frac{c}4\ln\sqrt{1+{t^2}/\epsilon^2}\simeq  \frac{c}4\ln t.
\ee
This prediction has been accurately checked against exact numerical computations for free fermionic models \cite{ds-11}.

\subsection{Decoupled finite interval}

A natural question is how all the results derived above
change when we cut and glue not half-line by a finite interval of length $\ell$.
It is straightforward to have a path integral for the density matrix: we
only need to have two pairs of slits for $-\infty$ to $-i\epsilon$ and from
$i\epsilon$ to $+i\infty$ at distance $\ell$. 
However, it becomes difficult to treat this case analytically.

In Ref. \cite{cc-07l} we preferred to treat the similar case in which an interval of length $\ell$ is embedded in 
half-line and start from the boundary. In this case we have only one pair of slits at $\ell$.  
The inverse of the conformal mapping from this geometry to the half-plane can be written analytically but cannot be 
inverted. Then, by making some physical motivated approximations on this  mapping between, it has been shown 
that for $t<2\ell$ the one-point function of a primary field should be given by \cite{cc-07l}
\be
\langle\Phi(i\tau)\rangle=
\left[\frac{\pi\epsilon}{4\ell}\frac{1}{t\sin(\pi t/2\ell)}\right]^{x_\Phi}\,. 
\ee 
This can be also used to derive the entanglement entropy between the initially decoupled interval and the rest of the 
system, obtaining \cite{cc-07l}
\be
S_A=\frac{c}{6}\ln\left(\frac{4\ell}{\pi\epsilon}t\sin(\pi t/2\ell)\right)+\tilde{c}'_1\,. 
\label{S2} 
\ee

In the case of an initially decoupled slit of length $\ell$ in an infinite chain Eq.~(\ref{S2}) is expected to be still valid 
with the replacements $2\ell\to\ell$ and $c/6\to c/3$ as follows from a simple analysis \cite{ekpp-07}. 
The validity of this has been carefully tested numerically for the XX chain in Ref.~\cite{ekpp-07},
finding very good agreement for all $t<\ell$. 
In Ref.~\cite{isl-09} a more complicated kind of defects has been investigated, and the results always agree with
Eq.~(\ref{S2}) when describing a conformal hamiltonian.

\subsection{Finite systems}

Dubail and St\'ephan \cite{ds-11} extended ingeniously the results of the previous subsections to the case in which at 
time $t=0$ we cut and glue two finite segments (of lengths $L_A$ and $L_B$  with $L=L_A+L_B$) 
in their respective ground states.
The starting point of the problem is indeed easy, one should just consider the geometry on the left of 
Fig.~\ref{map1},
but with the left and right sides having finite lengths $L_A$ and $L_B$ instead of being semi-infinite.  
Assuming that all the boundary conditions (i.e. at the four ends of the two segments) are all the same,
the physical quantities can be computed by mapping this geometry (called double pants in \cite{ds-11}) to the half-plane. 
However, finding out this mapping analytically is generically prohibitive, except in few easy circumstances. 

One of the relevant easy cases is when $L_A=L_B=L/2$, for which the mapping can be found analytically \cite{ds-11}.
We refer the readers interested in the details of calculations to the original reference \cite{ds-11} and we state only the final results. 
The entanglement entropy between the two initially disconnect halves evolves like
\be
S_A(t)=\frac{c}3 \ln \left| \frac{L}{\pi}\sin \frac{\pi t}{L}\right|+{\rm const}\,.
\label{Sfss}
\ee
Indeed this formula was previously guessed on the basis of numerical simulations \cite{isl-09,dir-11}.
It is also possible to treat the more general case when the bipartition does not coincide with the position of the quench,
as in Eq.~(\ref{s22}).  The final result is \cite{ds-11}
\be 
\fl
S_A(t)= \cases{
\frac{c}6 \ln L + {\rm const} & $0<t<\ell$ or $L-\ell<t<L$ ,\cr
\frac{c}6 \ln \frac{L^2}{\pi^2}\Big[\sin^2\frac{\pi t}{L}-\sin^2\frac{\pi \ell}L \Big] +{\rm const}' & $\ell<t<L-\ell$, } 
\ee
and $S_A(t)$ is periodic with period $L$. As a check, in the limit $t,\ell\ll L$ one recovers (\ref{s22}).

Along similar lines it is also possible to calculate the the return amplitude (\ref{LE}), obtaining \cite{ds-11}
\be
{\cal F}(t)=\frac{c}4 \ln \left| \frac{L}{\pi}\sin \frac{\pi t}{L}\right|+{\rm const}\,.
\label{Ffss}
\ee
Notice that all the above results are nothing but the those in an infinite geometry, where $t$ and $\ell$
have been replaced by their ``chord counterparts'' $\frac{L}{\pi} \sin \frac{\pi t}{L}$ and $\frac{L}{\pi} \sin \frac{\pi \ell}{L}$. 
As noticed in \cite{ds-11}, this simple relation is usually standard in CFT, but
 breaks down as soon as the quench is not located exactly in the middle of the strip. 

\begin{figure}[t]
\includegraphics[width=0.48\textwidth]{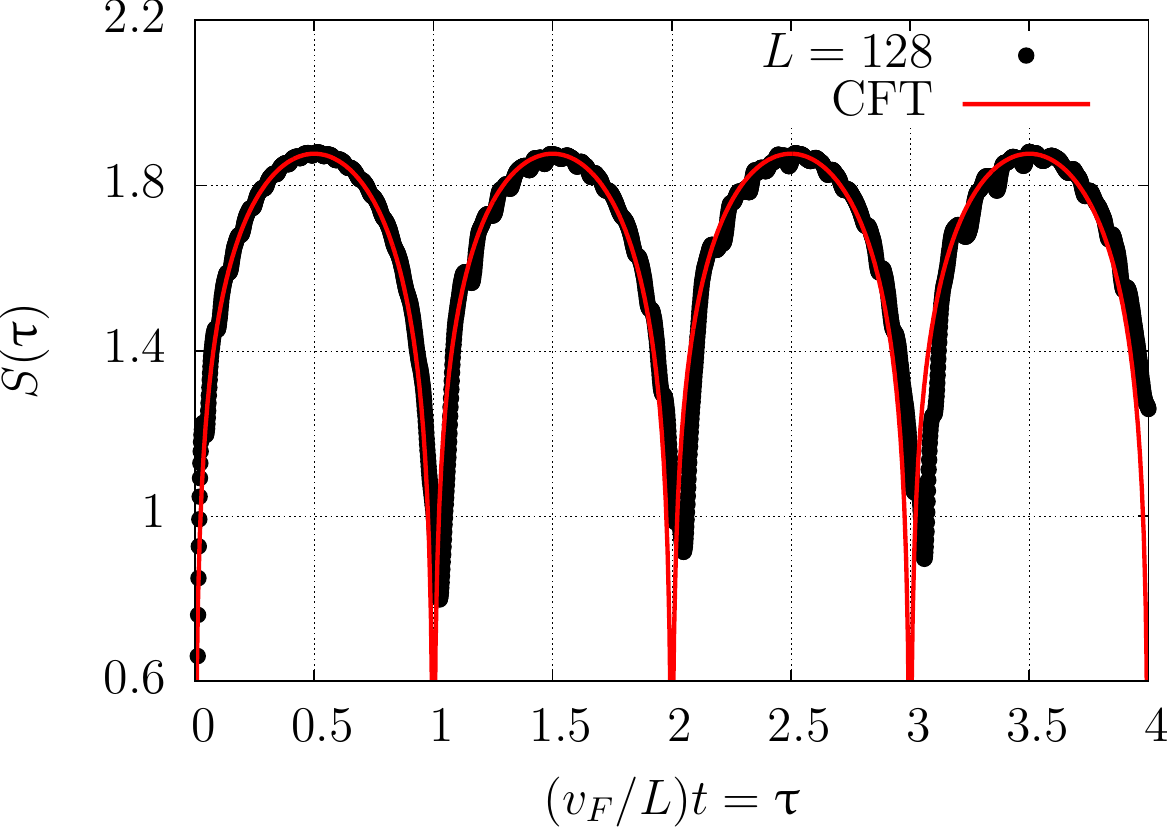}
\includegraphics[width=0.48\textwidth]{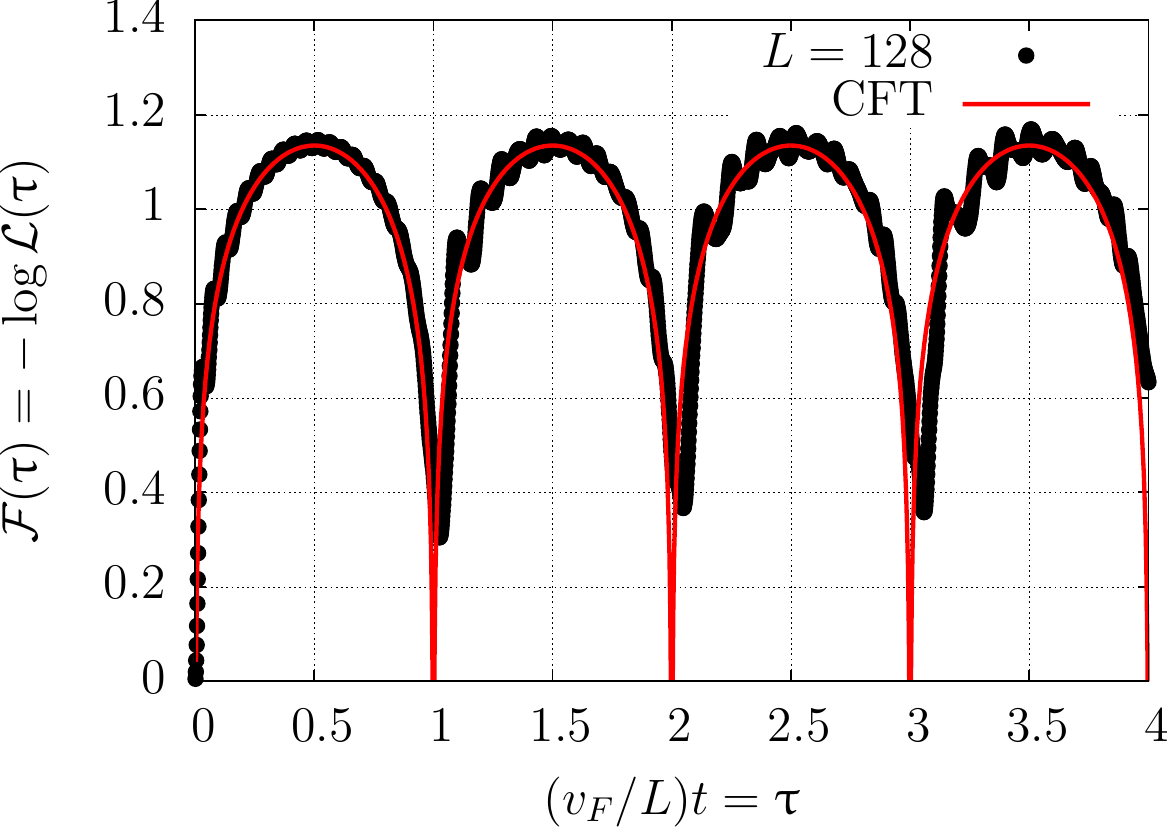}
\caption{
Entanglement entropy (left) and the logarithmic return amplitude (right) after a local quench in the middle of a system of size 
$L = 128$ for the XX spin-chain. Four oscillations of the finite-size system are clearly observed after the local quench. 
They are very well described by the CFT formulas (\ref{Sfss}) and (\ref{Ffss}). 
Reprinted with permission from \cite{ds-11}. 
} 
\label{ds-ee}
\end{figure}

Another important feature of the finite size systems is that all CFT results are periodic in time with period $L$.
This clearly reflects the commensurability of the spectrum of the CFT and the fundamental fact that 
for the particular type of quench we are looking at, only the identity and its descendants have non-zero overlaps
with the initial state \cite{ds-11}. 
When comparing with the results of lattice models whose scaling limit is described by a CFT, the commensurability 
of the spectrum is spoiled at order $L^{-2}$ and hence this periodic behaviour is expected to be visible for 
times such as $v t\ll L/a$ \cite{ds-11} (with $v$ being the sound velocity). 
Indeed in Fig.~\ref{ds-ee} we report the numerical results in \cite{ds-11} for both the entanglement entropy an the 
return amplitude which display this periodic behaviour. 
Indeed for local quenches, we expect only the lower energy part of the spectrum to be significantly excited and this is
 correctly described by CFT.  This is drastically different compared to global quenches in which high-energy excitations 
 with non-universal scaling behaviour are excited and lattice effects are much more important.

If one is interested in geometries where the cut is not in the middle of the chain ($L_A\neq L_B$), the problem is much more
complicated, and a full explicit solution is prohibitive.  
However, in \cite{ds-11} it was shown that for the return amplitude the difficulty can be circumvented by performing 
the analytic continuation semi-numerically and this is sufficient to provide the exact answer in the limit of $t,L_A,L_B\gg \epsilon$. 
For example, in the case when $L_B=\infty$ and $L_A$ arbitrary,  the final result can be written as \cite{ds-11}
\be
 \mathcal{F}(b(t)) =\frac{c}{6}\left(\Re e \left[ \frac{2}{1-b^2} \right]+\frac{3}{2}\log |b| -\log |b^2-1|\right).
 \label{ds-loch}
\ee
The parameter $b$ depends on time and $L_A$ and it is given by the solution of the transcendental equation
 \begin{equation}
 i v t+\epsilon=\tau(b)=\frac{2 L_A}{\pi} \left[\frac{2b}{1-b^2}+\log \left(\frac{b+1}{b-1}\right)\right].
\end{equation}
The above equation can be solved numerically for  $b(t)$ and successively can be injected in (\ref{ds-loch}) to have the 
time dependence of the return amplitude.
A similar analysis can be repeated for arbitrary $L_B$, but the final result is even more cumbersome to be written down 
and we refer the reader to Ref. \cite{ds-11}. 
The entanglement entropy turns out to be much more complicated in the asymmetric geometries.
In Ref. \cite{ds-11} only the case of aspect ratio $x=L_A/L=1/3$ was considered, but the final formula is too complicate to be reported here.
In Ref. \cite{ds-11} all these results have been compared to the numerical solution of free fermionic models, finding excellent 
agreement with the CFT predictions (in the proper regime of applicability) for all aspect ratios.

\subsection{Measuring entanglement using a local quench}

The possibility of measuring experimentally the entanglement entropy of a many-body quantum systems has longly been
a fascinating topic, made very difficult by the fact that it is an intrinsically non-local quantity. 
A few proposals in the literature are based on local quench protocols which we are going to briefly mention here, 
although the first entanglement measurement in cold atomic systems is based on different ideas \cite{meas}. 

The first idea due to Klich and Levitov \cite{kl-08} was to relate the entanglement after a local quench between
two half-chains to the distribution of non-interacting particles passing through  the contact between them. 
The main result of ~\cite{kl-08} is to establish a  relation between the cumulants $C_k$ of particle number fluctuations 
and the entanglement entropy of the two-halves of non-interacting fermions which reads
\be 
S_A=\sum_{k>0} {\alpha_k\over  k !}C_{k} ,\quad
\alpha_k= \Big\{
\begin{array}{cc}
    (2\pi)^k |B_{k}|, &  {\it k}\,\,{\rm even} \,,\\
  0,  &  {\it k}\,\,{\rm odd}\,,
\end{array} 
\label{Enoise} 
\ee 
where $B_m$ are Bernoulli numbers. The quantum noise generated by a switching on of the 
contact at $t=0$ is characterised by gaussian current fluctuations ($C_{m\neq 2}=0$),
with  variance growing with time as $C_2={1\over \pi^2}\ln t$. 
Combined with Eq.~(\ref{Enoise}) this gives entropy $S_A\sim (1/3)\ln t$
which is the  result after a local quantum quench (\ref{Slogt}).
As pointed out in \cite{hgf-09}, it is not clear how to generalise this idea to interacting models 
and to model without a conserved current. 
It is worth mention that relations similar to (\ref{Enoise}) have been derived also for the entanglement entropy of a 
finite interval in equilibrium \cite{fluc}.
 
A different way of measuring R\'enyi entanglement entropies of integer order has been proposed 
in Ref. \cite{c-11}. It is based on a particular local quench and it should be valid for an arbitrary system. 
The details of the proposal are probably too technical to be reported entirely here and so we describe only the main concepts. 
First of all it is convenient to introduce $n$ copies (replicas) of the original system and,
following \cite{hgkm-10}, define the swap operator $\Pi_n$ that maps each $j$-th replica to the $(j+1)$-th one in the subsystem 
$A$, while on $B$ acts as the identity. It is then clear \cite{zzf-00} that the expectation value of $\Pi_n$ on the 
ground state of the $n$-copy model with Hamiltonian $H$ (i.e. $ \ _H\langle 0| \Pi_n|0\rangle_H$) 
is the desired $n$-th moment of the reduced density matrix ${\rm Tr}\rho_A^n$.
Let us now define $H' = \Pi_n^{-1}H \Pi_n$. 
The hamiltonians $H'$ and $H$ are isospectral and $|0\rangle_{H'}=\Pi_n^{-1} |0\rangle_H$ is the ground state of $H'.$ 
Hence, thinking of $H\to H'$ as a quantum quench, the modulus squared of $ \ _H\langle 0| \Pi_n|0\rangle_H$ is 
the fidelity $P_0=  |_{H'}\langle 0| 0\rangle_H|^2$.
Obviously the fidelity itself is not directly measurable, but it is useful to think at the action of $\Pi_n$
on the replicated model. 
For example, for $n=2$, the action of $\Pi_n$ amounts to cut the two lines at the boundary points between $A$ and $B$
and to join at those points the two replicas (see \cite{c-11} for more details and for arbitrary order of the R\'enyi entropy $n$) 
and this is nothing but a particular local quench, usually called quantum switch.
The final argument of \cite{c-11} is that, in a system close to a quantum critical point described by a CFT, 
the measurable probability $P(E)$ of finding the system in a low-lying excited state of energy $E$ is given by $P_0$
(i.e. the desired R\'enyi entropy) times a calculable factor which depends on the relative geometry of $A$ and $B$.

Later, Abanin and Demler \cite{ad-12} pointed out that the overlap $P_0$ can be accessed by introducing 
a weak tunneling between the two states of the quantum switch, leading to an hybridisation of the two ground states.
Consequently, measuring Rabi oscillations between the two ground states would give direct access to $P_0$.
At the same time, in \cite{dpsz-12}  the action of $\Pi_n$  was related to the evolution of the parity number 
after a particular quench for two replicas of a bosonic chain. 

\section{Brief review of other quench protocols}
\label{sec:other}

In this last section we briefly review some generalisations of the already reported
quenches  which can be 
treated either with conformal field theory or with methods similar to those reviewed above. 
It is impossible to provide an account of all the aspects of quantum quenches in field theories and to their applications.
There are already excellent reviews on some aspects of the subject \cite{silva,efg-14,dkpr-15,ge-15} and a few more are present in this 
special volume. The goal of the following examples is just to show how the imaginary time formalism 
developed above can find further applications. 

Two different non-equilibrium situations in conformal field theory will be treated in other reviews in this 
volume and for this reasons will not be considered here. 
These are the quantum quench within the Luttinger model \cite{lutt} (a case which includes the possibility that also the initial state 
is the ground state of a CFT with algebraically decaying correlations) 
and the real time transport governed by CFT \cite{transp}.

\subsection{Inhomogeneous initial states}
\label{sec:inoh}

One natural question going beyond the results in the previous sections is to ask what happens when an initial 
state with an extensive excess of energy (as in a global quench) is not translational invariant (as in a local quench) 
and the post-quench Hamiltonian is always conformal invariant.
This is of particular relevance to understand several experimental situations, such as the presence of a harmonic 
confinement in cold atoms, voltage differences in quantum wires, and many more. 

\subsubsection{Smoothly varying initial state.}

In Ref. \cite{sc-08},  the case was considered when the initial state is of the form (\ref{tau0B}) but with 
$\tau_0$ being a slowly varying function of the space coordinate. 
This was motivated by the fact that being (in the proper limits) $\tau_0$ proportional to the correlation length
(inverse of the mass) in the initial state, such case should mimic an initial space varying mass.
A space dependent $\tau_0(x)$ corresponds in imaginary time to a 2D strip with variable width.  
This variable width strip can be transformed into a standard strip by a conformal transformation 
$w\to z=g(w)$. Then the transformation law of correlation functions of (primary) operators under such mappings 
will allow us to derive the corresponding expressions using already known results in the strip. 
This problem is mathematically well posed, but it usually difficult to solve analytically for an arbitrary $\tau_0(x)$.
There is however one case that can be treated in a general fashion, i.e. 
the limit in which the transformation is infinitesimal, that is $\tau_0(x)=\tau_0+h(x)$ varies only slightly in comparison 
with some average value, i.e.  $h(x)\ll \tau_0$.
It turns out that the behaviour of physical quantities for large times and separations is determined by
the asymptotic behaviour of the transformation for large $|z|$; 
since the initial distribution $h(x)$ is bounded, we can distinguish two important cases: either it tends to the same value as 
$x\to\pm\infty$ or to different values for each limit and the two can be understood just by considering  a bump or a step 
distribution localised at the origin.

A first non-trivial difference compared to the homogeneous case is that there is a net energy flow, given by the value 
of the $01$ component of the stress energy tensor. 
For infinitesimal $h(x)$ this can be computed exactly, obtaining \cite{sc-08} 
\be
\langle T_{01}(x,t)=\frac{c \pi}{96 \tau_0^3}(h(x+t)-h(x-t))\,,
\ee
i.e. the energy flows in a wave-like fashion. 
Correlation functions and entanglement entropy can be also calculated exactly. 
We refer the interested reader to Ref. \cite{sc-08} for all the results and   
we report only the final result for the entanglement entropy of the segment $A=[x_1,x_2]$
\bea
\fl S(x_1,x_2,t)&=& S_0(|x_1-x_2|,t)+ 
\frac{c}{12 \tau_0} (h(x_1-t)+h(x_1+t)+h(x_2-t)+h(x_2+t) )-\nonumber \\
\fl & -& \frac{c \pi}{24\tau_0^2} 
\Big[\theta(|x_1-x_2|-2t) \Big( \int_{x_1-t}^{x_1+t} + \int_{x_2-t}^{x_2+t} ds h(s)\Big) + \nonumber \\ \fl &&
+ \theta(2t-|x_1-x_2|) \Big( \int_{x_1-t}^{x_2-t} + \int_{x_1+t}^{x_2+t} ds h(s)\Big) \Big]\,,
\eea
where $S_0$ is homogenous result (\ref{SAt2}).

\subsubsection{Domain-wall initial state without conservation laws.}

A different inhomogeneous situation has been considered in \cite{chl-08}.
In this case the initial state is of the form (\ref{tau0B}) to the left and to the right of a given point 
(let us say $0$), but the two halves correspond to different boundary conditions with a changing operator between them.
For example, for the simple case of the Ising CFT, this includes the case where all the spins on the left are aligned in one 
direction and all the spins on the right are in the opposite direction. 
The resulting time evolution is easily understood on general considerations: 
for $|x|>t$ the two halves evolve independently since  one does not causally feel the presence of the other, 
while inside the light-cone $|x|<t$ there is a non trivial evolution with spin transport and complicated correlation functions. 
Many results can be derived by analytically continuing known correlation functions for strips with different boundary conditions,
but again for lack of space we refer the interested reader to the original reference \cite{chl-08}. 

\subsubsection{Domain-wall initial state with a conserved $U(1)$ current.}

We now discuss a quench which has features  between a global and a local one.
This can be realised in models that possess a $U(1)$ symmetry such us XXZ spin-chains or models of spinless fermions. 
Let us for example take two uniformly filled half-chains of spinless fermions in their respective ground-states, 
but with a different densities. At time $t=0$ the two are connected as in a local quench to form an infinite chain. 
There are clearly particle and energy currents flowing from one edge to the other and the dynamics take place only 
inside a light-cone propagating from the joining point, because outside of it the system is locally in an eigenstate. 
(For equal filling, this is nothing but the local quench of the previous section without energy current.)
Another interesting and slightly simpler limit is the case in which half-chain is prepared with filling $1$ and the 
other half $0$, i.e. in spin language there is a domain wall between two fully and oppositely polarised half-chains. 
This quench protocol is studied since long time and numerous numerical and analytic results are available 
from different techniques \cite{arrs-99,er-13,akr-08,mpc-10,lm-10,sm-13,ah-14,vsdm-15}, especially in the free fermions case  
when calculations are rather elementary (see \em e.g. \em\cite{vsdm-15}).

\begin{figure}[t]
\includegraphics[width=\textwidth]{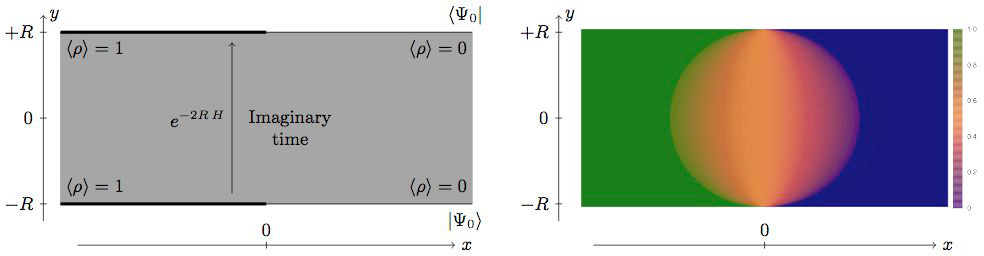}
	\caption{Left: Imaginary time representation of the domain-wall initial state quench.
	Right: Density profile in the strip geometry measured numerically for $R = 128$. 
	Reprinted with permission from \cite{adsv-16}.
	}
	\label{fig:param2}
\end{figure} 

Although the large number of works on the subject, a field theoretical description of the problem in imaginary time 
has been only recently proposed for the case of free fermions \cite{adsv-16}. 
Following the logic introduced for global quenches, it is rather natural to expect that the imaginary time version of this 
quench problem is given by the strip geometry depicted on the left of Fig.~\ref{fig:param2}, in which the two boundaries are 
domain-wall states. Unfortunately the analogy with global quenches finishes here because those states have nothing to 
do with conformally invariant boundary states.
The idea of Ref. \cite{adsv-16} proceeds as follows.
At imaginary time $y=-R$ the degrees of freedom are frozen for any $x$. 
For larger imaginary times, the particles are released and can hop to  left or to the right. 
So, as soon as $y > -R$, a small region around $x = 0$ appears, where 
there are density fluctuations, with non trivial correlations. 
The strip is symmetric for reflection respect to the real axis. 
In the scaling limit $x,y,R\to\infty$ with $x/R$ and $y/R$ constant, 
the exact shape of the fluctuating region is the disc of radius R centred at the origin \cite{adsv-16}, see 
Fig.~\ref{fig:param2}.

The main problem  now is to understand the field theory inside this disc.
Clearly, it should be massless.  Presumably there should exists also a single theory both in the interior and in the exterior of 
the disc, with a non-vanishing mass only outside the disc.
For the case of a free fermion, this field theory must be a massless Dirac fermion (in more general cases
 we expect a compactified boson/Luttinger model), but the action is non-trivial in order to describe a fluctuating region. 
The non-triviality of the action comes from having a non flat measure inside the fluctuating region, a metric that can be 
fixed by taking the continuing limit of the hopping model, see for details \cite{adsv-16}.
 
\subsubsection{More general local quenches in CFT.}

The cut and glue protocol is only one of the many ways to generate a local quench. 
More general situations in spin-chains or gas models are such that a local density or phase perturbation
is generated locally and then the following time-evolution is studied, see \em e.g. \em\cite{kk-08,gr-12,fs-13,la-14}.
In general it is not simple to model these situations in field theories.
In Ref. \cite{ntt-14} it has been considered the case in which this local perturbation is obtained by acting 
with a given conformal local operator at some point. 
It turns out that the evolution of the entanglement entropy depends on the considered operator in a calculable manner 
(see for more details also \cite{cnt-14}). 

Insertions of relevant operators have also been used in \cite{e-16} for a very different kind of global quench. 
There, it was proposed a particular boundary state on the strip of width $2\tau_0$ to mimic 
the split momentum quench in the famous quantum Newton cradle \cite{kww-06}. 
A pair of chiral and antichiral vertex operators have been 
inserted on the boundary of a compactified free boson with Dirichlet conditions. 
The opposite-chirality vertex operators act to excite the ground state in analogy to the experimental setup of two excited 
oppositely moving bosonic clouds. 
The results found with this approach qualitative resemble the momentum distribution function measured in the experiment.

\subsection{Higher dimensions.}

The imaginary time formalism used extensively for (1+1) dimensional CFT in Sec.~\ref{sec:global} is not limited 
to one dimension. 
As already exploited in \cite{cc-06}, the real-time evolution of a $d$-dimensional quantum system can always be interpreted 
as a boundary problem in $d+1$ dimensions in imaginary time. 
In this case, instead of a strip we have a slab of width $2\tau_0$. 
Unfortunately, this is not generically very insightful because it is very complicated to solve field theories in confined 
geometries in arbitrary dimensions, even conformal invariant ones, since we do not have at disposal the powerful tools 
of conformal mapping (aka the conformal group is finite dimensional). 

However, mean field models can be exactly solved. 
In \cite{cc-07} the case of a critical mean-field evolution was considered and one of the most remarkable results was 
to show that the order parameter after the quench can display persistent non harmonic oscillations depending on the 
initial state. 
This finding was generalised to arbitrary massive post-quench mean field hamiltonian in \cite{cg-11}.
This general solution allowed for the reconstruction of a non-equilibrium mean-field dynamical phase diagram 
(in function of pre- and post-quench masses) which closely resembles the results obtained in real time 
for specific bosonic and fermionic mean field models \cite{sb-10,hub,bl-06}. 
Taking into account fluctuations beyond mean field is a very difficult problem which is still at a very embryonic 
level, see \em e.g. \em\cite{sc-10,msf-14,ctgm-15,mcmg-15,lm-16} as relevant manuscripts on the subject. 

Finally it must be mentioned once again that in higher dimensions many results are available from holographic methods 
(see \em e.g. \em\cite{hol1,hol2,at-12,bmn-13,nnt-13,rrv-16}) and free field theories \cite{cc-07,dgm-14,sm-16}.

\section*{Acknowledgments}
We are indebted to many colleagues for numerous discussions and collaborations on quantum quenches 
during the last decade, in particular we need to thank Vincenzo Alba, J.-S. Caux, Mario Collura, Fabian Essler, 
Maurizio Fagotti, Rosario Fazio, Andrea Gambassi, Robert Konik, 
Marton Kormos, Giuseppe Mussardo, Lorenzo Piroli, Alessandro Silva, Spyros Sotiriadis.
PC acknowledges the financial support by the ERC under Starting Grant 279391 EDEQS,
and JC that of the Miller Institute at UC Berkeley.

\end{document}